\documentclass[10pt, twocolumn,prd,aps,floatfix,preprintnumbers,a4paper,nofootinbib,superscriptaddress]{revtex4-2}
\usepackage[utf8]{inputenc}
\UseRawInputEncoding
\usepackage{amsmath,amssymb}
\usepackage{amsfonts}
\usepackage{bm}
\usepackage{courier}
\usepackage{graphics}
\usepackage{comment}
\usepackage{amsmath}
\usepackage{amssymb}
\usepackage{xcolor}
\usepackage[normalem]{ulem} 
\usepackage[mathscr]{euscript}
\usepackage{acronym}
\usepackage{multirow}
\usepackage{graphicx}

\usepackage{siunitx}
\usepackage[justification=justified,singlelinecheck=false]{caption}
  
\usepackage{subcaption}  
\usepackage[dvipsnames,table,xcdraw]{xcolor}
\usepackage[colorlinks, pdfborder={0 0 0}]{hyperref} 
\usepackage[nameinlink,capitalise]{cleveref}
\usepackage{tensor}
\usepackage{verbatim}
\usepackage{wrapfig}
\usepackage{soul} 

\usepackage{tikz}
\usepackage{tcolorbox}
\usetikzlibrary{decorations.pathmorphing, patterns, calc, arrows.meta, fadings}

\definecolor{inspiral}{RGB}{180,220,255}
\definecolor{inspiralborder}{RGB}{60,130,210}
\definecolor{plunge}{RGB}{255,180,180}      
\definecolor{farzone}{RGB}{230,245,225}
\definecolor{bhcolor}{RGB}{10,10,10}
\definecolor{cocolor}{RGB}{80,80,80}
\definecolor{orbitcolor}{RGB}{40,100,200}

\graphicspath{{fig}}

\newcommand{\uib}{
Departament de F\'isica, Universitat de les Illes Balears, IAC3 – IEEC, Crta. Valldemossa km 7.5,
E-07122 Palma, Spain}

\usepackage{amssymb}

\begin{document}
\title{Impact of the axion-like self-interactions in gravitational atoms for LISA}
\author{Samuel G\'omez G\'omez}
\email{samuel.gomez@uib.cat}
\affiliation{\uib}

\author{Xisco Jimenez Forteza}
\email{f.jimenez@uib.es}
\affiliation{\uib}

\author{Carlos Palenzuela}
\email{carlos.palenzuela@uib.es}
\affiliation{\uib}

\begin{abstract}
Ultralight bosons with self-interactions, such as axion-like particles, can form astrophysical Bose--Einstein condensates around stars or compact objects, often referred to as gravitational atoms. In this work, we adopt a recently proposed dynamical formation mechanism for these halos and estimate their impact on extreme- and intermediate-mass-ratio inspirals when present around the primary black hole. We show that, for signal-to-noise ratios $\lesssim 100$, LISA can distinguish gravitational waveforms from binaries embedded in such halo overdensities. Our analysis indicates that LISA can probe boson masses $m_\mathrm{dm}\sim10^{-17}$--$10^{-15}\,\mathrm{eV}$ and decay constants $f_a\sim3 \times 10^{10}$--$6 \times 10^{12}\,\mathrm{GeV}$ using binaries with total masses $M\sim10^4$--$10^5\,M_\odot$, assuming conservative DM densities consistent with the central values of Navarro--Frenk--White profiles. Allowing for higher background densities and different extreme-mass-ratio configurations further extends the accessible parameter space. Moreover, we find that for a binary configuration with $M\sim10^4M_{\odot}$, $\rho_{\rm dm} = 10^4\,\mathrm{GeV/cm^3}$, and signal to noise ratio $\text{SNR} \sim 20$, a particle mass of $m_{dm} = 3.2 \cdot 10^{-15}$ eV and decay constant of $f_a = 1.6 \cdot 10^{11}$ GeV maximize the dephasing due to dynamical friction, enabling the recovery of the particle parameters at the percent level. These results demonstrate that LISA can place constraints on axion-like particle masses and self-interactions without requiring additional couplings to Standard Model fields.
\end{abstract}

\maketitle
\section{Introduction}

Gravitational-wave (GW) observations have revolutionized modern astrophysics through the direct detection of compact binary coalescences across four observing runs of the LIGO--Virgo--KAGRA (LVK) network, opening an entirely new observational window onto the Universe and enabling precision studies of compact objects~\cite{Abbott_2019, Abbott_2021, Abbott_2023, theligoscientificcollaboration2025gwtc40updatinggravitationalwavetransient}.  Next-generation detectors, including the Laser Interferometer Space Antenna (LISA)~\cite{colpi2024lisadefinitionstudyreport} and the Einstein Telescope~\cite{Abac_2026, Branchesi_2023, Maggiore_2020}, are expected to dramatically extend our reach in frequency and sensitivity. GW observations open new opportunities to address important questions in cosmology and astrophysics, and potentially in fundamental physics \cite{Arun_2022, Askar_2019, Yunes:2013dva, Cardoso_2016, berti2025blackholespectroscopytheory}. Among these, one of the most open problems is understanding the nature and distribution of dark matter (DM)~\cite{Bertone_2020, Brito_2020, Cardoso_2018, Brito_2017, Macedo_2013, Eda_2013, Eda_2015, Kavanagh_2020, Duque_2024, Sasaki_2018, Bartolo_2019}.

The possible microscopic nature of DM is constrained by its quantum spin, allowing for scenarios in which it consists of bosons, fermions, or a mixture of both. In this paper, we focus on the possibility that DM is composed of light, or ultralight bosonic particles, which are a broad class of hypothetical integer spin particles with masses between $10^{-24}$ to $1$ eV, that have recently attracted the attention of a large part of the community, especially after the WIMP parameter space has become increasingly constrained \cite{Arcadi_2025}. Moreover, although the $\Lambda$CDM model exhibits outstanding agreement with current cosmological observations, it remains very much unconstrained for subgalactic scales (typically below $k \lesssim 0.1 \mathrm{kpc}^{-1}$), and light bosons can address the small scale challenges of $\Lambda$CDM by offering a rich phenomenology which can be tested with astrophysical observations \cite{Ferreira_2021}. 

The most familiar example of a light boson coming from well-motivated extensions to the Standard Model is a pseudoscalar known as the axion \cite{PhysRevD.95.043541}, such as the QCD axion, a pseudo-Goldstone boson originally proposed as a solution to the Strong-CP problem \cite{PhysRevLett.38.1440, PhysRevLett.40.223, PhysRevLett.40.279}. More generally, light axions known as Axion Like Particles (ALPs) are also a generic prediction of different extensions to the Standard Model or are produced by the compactifications of extra dimensions, such as the moduli fields appearing in the different formulations of String Theory \cite{Arvanitaki_2010, Svrcek_2006}. In general, ALPs can have a wide range of masses and interactions depending on their decay constant $f_a$, which can have values anywhere from the weak scale ($\sim$ TeV) to the GUT or Planck scale, and their couplings to photons, other matter fields, and themselves, which are highly model-dependent.

Binary black holes (BBHs) with very unequal masses, such as extreme mass ratio inspirals (EMRIs) or intermediate mass ratio inspirals (IMRIs), have been proposed as promising scenarios for testing the presence of DM overdensities. The presence of a DM cloud around a central BH in one of these systems, can induce environmental effects on the smaller companion \cite{Barausse_2014, Barausse_2015,CanevaSantoro:2023aol}. This has been widely studied in the context of DM spikes and minispikes \cite{Eda_2013, Eda_2015, Zhao_2005, cole2022disksspikescloudsdistinguishing} without relying on microphysical assumptions about the DM. These systems are primary targets of LISA, the upcoming space-based GW detector, making it uniquely suited to detect subtle environmental imprints on the waveform. LISA is sensitive in the frequency band $f \sim 10^{-4}$--$10^{-1}$ Hz, with peak sensitivity  around $f \sim 3 \times 10^{-3}$ Hz. EMRIs, with mass ratios $q = m/M_{BH} \sim 10^{-4}$- $10^{-7}$,  emit in the $10^{-3}$--$10^{-1}$ Hz range and accumulate SNR coherently over $\sim 10^4$--$10^5$ cycles, achieving typical SNRs of $\sim 20$--$100$ at $z \lesssim 0.5$. IMRIs, with mass ratios $q \sim 10^{-2}$--$10^{-4}$, emit in the $10^{-3}$--$10^{-2}$ Hz range with SNRs reaching $\sim 10^2$--$10^4$ for favorable configurations at $z \lesssim 1$. 

Bosonic DM overdensities of light bosons or ALPs, are known to form around rotating Kerr BHs, exponentially growing by extracting energy and angular momentum in the ergoregion, in a process known as a superradiant instability \cite{Brito_2020,  Press:1972zz, Baumann_2022, Tomaselli_2023}, forming macroscopic quantum states known as gravitational atoms (GA). Allowing scalar bosons to have quartic self-interactions, it has recently been shown in \cite{budker2023genericformationmechanismultralight} that GAs can be formed and stabilized in gravitational potentials, while a relativistic stationary framework is introduced in~\cite{Chia:2022udn}. More recently, in~\cite{Aurrekoetxea_2024}, the full non-linear case has been explored in the context of numerical relativity, demonstrating that BH spacetimes can stabilize the cloud growth and lead to a stationary state.

Unlike previous studies based on phenomenological DM profiles of superradiantly generated clouds, the model in \cite{budker2023genericformationmechanismultralight} which we consider, the GA grows through a self-interaction--driven capture mechanism, which provides a direct link between the microphysical properties of the bosonic field and its macroscopic gravitational imprint. In this framework, the halo forms dynamically from the ambient DM distribution and evolves over time, rather than being assumed as a pre-existing static configuration. We model the time-dependent growth of the halo around the primary BH and compute its impact on the orbital evolution of EMRIs and IMRIs, including the effects of dynamical friction and accretion. These environmental effects modify the GW phase evolution, leading to a characteristic dephasing in the LISA band. We quantify the detectability of this signal and identify the regions of parameter space in which such effects can be probed for hosting BHs with masses $M_\odot \in \left[10^4,10^7\right]$. 

Recent works such as \cite{Boudon_2024, banik2025bosonstarshostingblack, Chavanis:2019bnu} have studied these configurations for specific GA overdensities, finding central densities of order $10^{21}\,\mathrm{GeV/cm^3}$ or higher for a central BH mass of $10^5\,M_\odot$. In these scenarios, the gravitational potential generated by the boson cloud must be taken into account, particularly when the cloud mass becomes comparable to that of the host BH ~\cite{banik2025bosonstarshostingblack}. The model we use describes a dynamical GA formation seeded by a realistic local galactic DM background density $\rho_{dm}\sim 10^3-10^4 GeV/cm^3$, rather than assuming a fixed overdensity as an initial condition. In particular, the maximum halo mass we obtain is of order $10^{-6}\,M_{\rm BH}$~\cite{budker2023genericformationmechanismultralight}, which justifies neglecting its self-gravity and treating the gravitational potential as dominated by the host BH. Within this framework, the maximum GA densities achievable near the BH horizon, for the largest BH mass we study $M_{BH} = 10^7M_{\odot}$, are $\rho_{GA} \sim 10^{20}\,\mathrm{GeV/cm^3}$, and only for background densities as large as $\rho_{dm} \sim 10^8\,\mathrm{GeV/cm^3}$, which are beyond realistic galactic environments. Moreover, even allowing for such extreme background densities, the companion would not probe these peak values for most of the binary evolution, since the GA density profile decreases exponentially with radius, so that the orbit-averaged density experienced by the secondary is substantially lower than the central value.

To date, there is no direct evidence for the presence of
environments in current GW detector data. The most stringent constraints come from \cite{CanevaSantoro:2023aol}, imposing the bound $\rho \lesssim 21\,\mathrm{g\,cm^{-3}}$ on the matter density. More recently, similar constraints on scalar-field environments around compact binaries have been obtained from LVK data using a semi-analytic waveform model based on the nonrelativistic Schr\"{o}dinger–Poisson description of scalar fields around BH binaries~\cite{Roy_2026}. These constraints apply jointly to central halo densities and particle masses, assuming superradiant priors. We note that if we apply the model of \cite{budker2023genericformationmechanismultralight} to BHs
with masses within the LVK range
$\sim (10\text{--}100)\,M_{\odot}$, we see that in order to
obtain GA radii of the order of the orbital
length scale at LIGO frequencies, we require ALP masses of
$\sim (10^{-12}\text{--}10^{-13})\,\mathrm{eV}$, consistent
with the fact that the de Broglie wavelengths should be of the order of the gravitational radii of the BHs. However, the densities involved in this work are $\sim 20$ orders of
magnitude larger than those reachable within the formation
timescales of the mechanism we employ.

Throughout this work, we work in Planck units ($G = c = \hbar=1$) and adopt the spacetime signature $[-1, 1, 1, 1]$, $M_{BH}$ is the mass of the primary central BH in the considered binaries, where the GA is forming; $m$ is the mass of the secondary BH; the chirp mass is defined as $\mathcal{M}_c = (M_{BH} m)^{3/5}/(M_{BH} + m)^{1/5}$, $q=m/M_{BH}$ is the mass ratio, $M$ is the total mass,  $\rho_{dm}$ is the background DM density; $m_{dm}$ is the ALP mass; and $f_a$ is the ALP decay constant.

\section{Formation mechanism for gravitational atoms}

If the ALP is an axion with mass arising from weakly-coupled instantons, it would be described by a periodic potential with the form \cite{Marsh_2016,budker2023genericformationmechanismultralight}
\begin{equation}
\label{eq:axion_potential}
V(\phi)=-m_{dm}^2 f_a^2 \cos \left(\frac{\phi}{f_a}
\right).
\end{equation}
where $m_{dm}$ is the ALP mass and $f_a$ is the ALP decay constant. Expanding the cosine
\begin{equation}
\label{eq:potential}
V(\phi)=\frac{1}{2} m_{dm}^2 \phi^2+\frac{\lambda}{4!} \phi^4+\ldots,
\end{equation}
we recover the case of a scalar field with quartic self-interactions. The sign of $\lambda$ determines whether the quartic self-interactions are attractive (negative) or repulsive (positive). It is useful to express $\lambda$ in terms of a dimensionful coupling $g$ as
\begin{equation}
\label{eq:coupling}
\lambda \equiv 8 g m_{dm}^2, 
\end{equation}
with $|g|=1/\left(8f_a^2\right)$ and $|\lambda| =  (m_{dm}/ f_a)^2$. We conduct our analysis using the decay constant $f_{\mathrm{a}}$, but the results can be equivalently expressed in terms of the quartic coupling $\lambda$, which provides a more model-independent description of the self-interactions.

Although some of these particles could couple weakly to other interactions, such as the QCD axion, which can couple to electromagnetism changing the QED lagrangian via a term $\mathcal{L}_{\text {int }}=g_{\phi \gamma \gamma} \phi F_{\mu \nu} \tilde{F}^{\mu \nu}$, where $F_{\mu \nu}$ is the electromagnetic strength tensor, $\tilde{F}^{\mu \nu}$ its dual and $g_{a \gamma \gamma}$ is the axion-photon coupling, this is not always guaranteed for ALPs. Moreover, there is no direct observational evidence that DM couples to other Standard Model sectors. We thus assume that this bosonic DM couples to gravity, has weak self interactions coming from the potential \eqref{eq:axion_potential}, and that other possible couplings are very small.

In the more general case we work with a scalar field on a generic spacetime background, whose action is given by
\begin{equation}
\begin{aligned}
S = \int d^4x \sqrt{-g}\Bigg[
&\frac{1}{16 \pi} R
- \frac{1}{2} g^{\mu\nu} \nabla_\mu\phi \nabla_\nu\phi \\
&- \frac{1}{2} m_{dm}^2 \phi^2
- \frac{\lambda}{4!}\phi^4
\Bigg]
\end{aligned}
\label{eq:action}
\end{equation}
Varying this functional with respect to the metric gives the Einstein field equations coupled to a scalar field 
\begin{equation}
G_{\mu \nu}=8\pi T_{\mu \nu} ,
\label{eq:Einstein_eqs}
\end{equation}
with the energy-momentum tensor
\begin{equation}
\begin{aligned}
T_{\mu \nu}
&= \nabla_\mu \phi \nabla_\nu \phi
- g_{\mu \nu} \left( \frac{1}{2} g^{\lambda \sigma} \nabla_\lambda\phi \nabla_\sigma\phi \right. \\
&\quad \left. + \frac{1}{2} m_{dm}^2 \phi^2 + \frac{\lambda}{4!}\phi^4 \right)
\label{eq:EM_tensor}
\end{aligned}
\end{equation}
Varying this action with respect to the axion field $\phi$, we obtain the Klein-Gordon equation in a generic spacetime 
\begin{equation}
(\square_g + m^2_{dm})\phi= \frac{\lambda}{3!} \phi^3,
\label{eq:KG}
\end{equation}
where $\square_g \phi=\nabla^\mu \nabla_\mu \phi$ is the curved d'Alembertian. This field equation of motion can be interpreted as arising from a mean-field approximation, which remains valid provided the field amplitude is small compared to the characteristic field range, its de Broglie wavelength is large relative to astrophysical length scales, and the field has a high occupation number.

For individual, astrophysical BHs, the full description would involve solving Eq. \eqref{eq:KG} for a BH metric. This is what has usually been done numerically in a great part of the literature to study the superradiance instability of vector bosons around Kerr BHs \cite{Pere_iguez_2024, Degollado_2018}. 

\begin{figure}
    \centering   \includegraphics[width=1\linewidth]{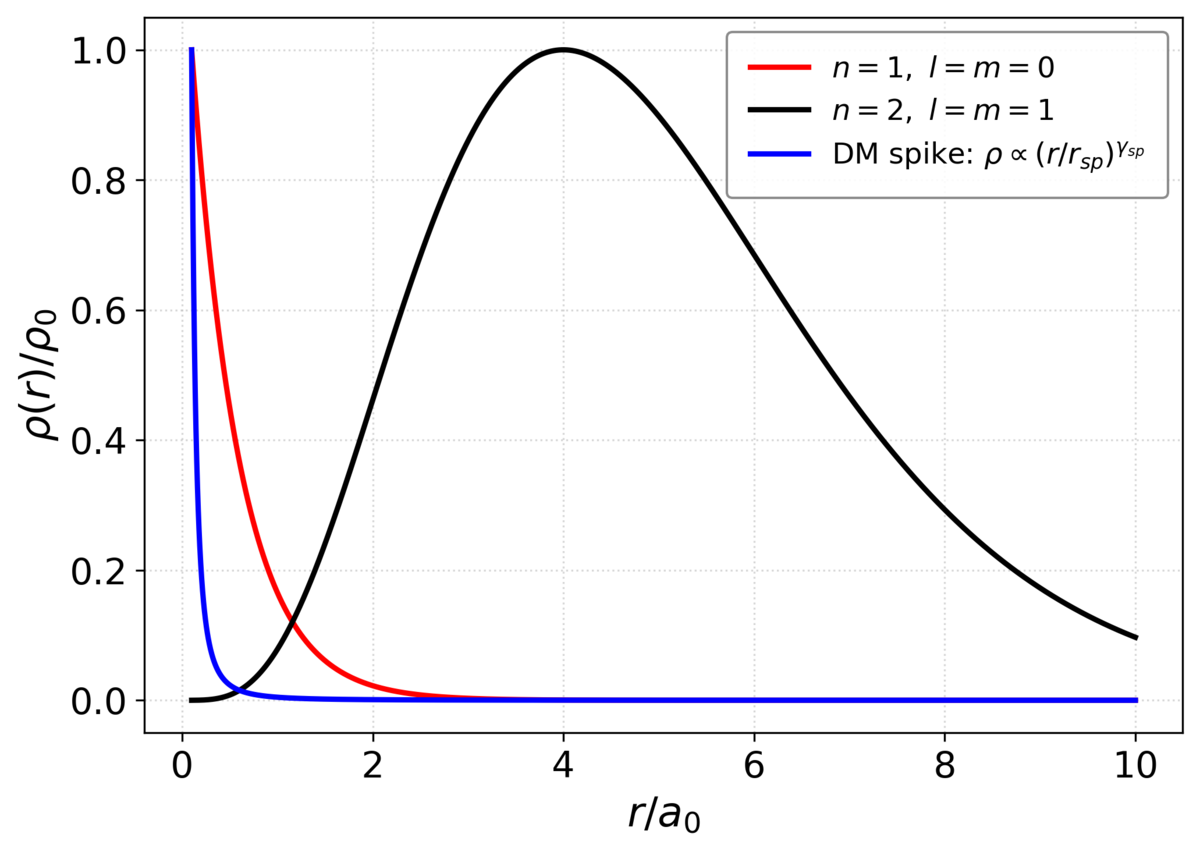}
    \caption{Density profiles for DM overdensities. The GA fundamental state we are considering, $n=1,\; l=m=0$, decays monotonically and exponentially from the origin. The superradiant, $n=2,\; l=m=1$ profile, vanishes at the center, rises to a maximum at finite radius $r \sim \mathcal{O}(a_0)$, and then decays smoothly at large radii. In contrast, the DM spike is characterized by a pronounced power-law cusp near the origin, followed by a rapid decline, becoming negligible at large radii. Different DM profiles present in EMRIs will produce different environmental imprints in the GW waveforms.}
    \label{fig:density_profiles}
\end{figure}

Fig.~\ref{fig:density_profiles} shows the normalized density profiles for the fundamental state of a GA ($n=1$), its first excited state ($n=2$) formed by superradiance and a radial profile for a DM spike \cite{Gondolo_1999}. The $n=1,\; l=m=0$ profile is centrally peaked and decays monotonically and exponentially from the origin. The $n=2,\; l=m=1$ profile instead vanishes at the center, rises to a maximum at finite radius $r \sim \mathcal{O}(a_0)$, and then decays smoothly at large radii. In contrast, the DM spike exhibits a steep power-law cusp near the origin, sharply decreasing and becoming negligible at larger distances. Different DM profiles of halos hosting central massive BHs in EMRIs will produce different environmental imprints in the GW waveforms, potentially allowing us to differentiate between DM models \cite{Cole_2023}. 
Other different DM profiles around BHs have been used in the literature, such as NFW galactic-like profiles \cite{Navarro_1997}, Hernquist \cite{1965TrAlm...5...87E} or Einasto \cite{1990ApJ...356..359H} profiles. Most of these profiles are all phenomenological in nature, motivated by N-body simulations or observations, and derived using Newtonian gravity alone, with the possible exception of superradiant clouds. Making the same assumption, by working in the Newtonian regime, with $g_{00}=-(1+2\Phi), g_{ij}=(1-2\Phi)\delta_{ij}$  where $\Phi$ is the Newtonian gravitational potential, and in terms of a nonrelativistic field $\psi$ defined as
\begin{equation}
\phi \equiv \frac{1}{\sqrt{2 m_{dm}}}\left(\psi e^{-i m_{dm} t}+\psi^* e^{i m_{dm} t}\right),
\label{eq:ansatz}
\end{equation}
in the limit where $\psi$ is slowly varying, $\dot{\psi} \ll m \psi$, this ansatz in Eq. \eqref{eq:KG} leads to the so-called Gross-Pitaevskii equation \cite{Gross:1961kqh, Pitaevskii_1996}.
\begin{equation}
    \left(i \partial_t+\frac{\nabla^2}{2m_{dm}}-m_{dm} \Phi\right) \psi=g|\psi|^2 \psi.
\label{eq:GP}
\end{equation}
There are two interesting choices for the gravitational potential $\Phi$. One is to take the potential generated by the bosons themselves, such that we would obtain a self-gravitating object known as a boson star, which are one of the most popular exotic compact objects that have been proposed as BH mimickers \cite{Liebling_2023, Bezares_2017, Bezares_2018}. The other possibility are GAs, which emerge by choosing the Newtonian potential generated by an external object, i.e. $\Phi=V_{\text {ex }} \equiv$ $-M_{BH} / r$. Choosing the latter option, the GP Eq. \eqref{eq:GP} reduces to
\begin{equation}
    \left(i \partial_t+\frac{\nabla^2}{2 m_{dm}}+\frac{\alpha}{r}\right) \psi=g|\psi|^2 \psi,
\label{eq:GA}
\end{equation}
with two couplings: the self interaction coupling $g$ and a gravitational coupling $\alpha \equiv M_{BH} m_{dm}$, which measures the strength with which the particles are pulled towards the potential well.

In the absence of self-interactions ($g=0$), the previous equation reduces to the Schr\"{o}dinger equation with a $1/r$ potential, being thus identical to the equation whose solutions are the wave functions of the hydrogen atom, but instead on depending on the electron mass $m_e$ and the fine structure constant ($\alpha=\frac{e^2}{4 \pi \varepsilon_0}$ where $e$ is the electron charge and $\varepsilon_0$ is the electric constant), it depends on the axion mass $m_{dm}$ and the mass of the astrophysical object $M_{BH}$. These bound states have a Bohr radius of the form
\begin{equation}
    a_0 = (m_{dm} \alpha)^{-1}=\frac{1}{M_{BH} m_{dm}^2}.
\label{eq:bohr}
\end{equation}
Bound solutions ($E<0$) correspond to the well-known hydrogenic wave functions, with ground state $\psi^{(0)} \propto e^{-r/a_0}$ and density profile $\rho \propto e^{-2r/a_0}$. 
\begin{figure}[htb]
\centering
\begin{tikzpicture}[scale=1.4,
                    BHbig/.style={ball color=black},
                    BHsmall/.style={ball color=black},
                    DMfill/.style={blue,opacity=#1,draw=none}]

  \def\a{1.0}     
  \def\RDM{1.7}   

  \foreach \i in {1,...,20} {
    \pgfmathsetmacro{\radius}{\i*\RDM*\a/20}
    \pgfmathsetmacro{\opacity}{0.9*exp(-2*\radius/\a)}
    \fill[DMfill={\opacity}] (0,0) circle (\radius);
  }

  \shade[BHbig] (0,0) circle (0.30);
  \node[below=2pt] at (0,-0.25) {\small $M_{BH}$};

  \draw[->, red]

  node[pos=1, scale=0.75, above=5pt, right=22pt, red] {$a_0$};  
  \pgfmathsetmacro{\radio}{0.8*\a};
  \draw[red, dashed] (0,0) circle (\radio);

\end{tikzpicture}

\caption{GA sourced by a BH of mass $M_{BH}$ and a background DM density $\rho_{dm}$, creating a bosonic cloud overdensity with density profile $\rho(r)\propto e^{-2r/a_0}$, with $\rho \gg \rho_{dm}$. Its Bohr radius $a_0 = 1/(M_{BH} m_{dm^2})$ is marked in red. }
\label{fig:GA}
\end{figure}
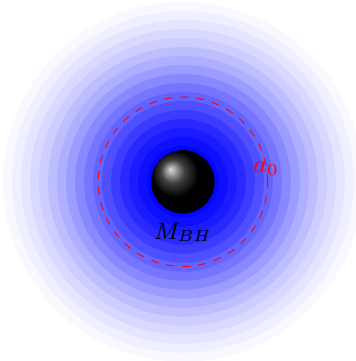
Unbound solutions or scattering states ($E>0$), correspond to traveling waves which form a continuum of states, $\Psi \propto \psi_{\mathbf{k}}$, are parametrized by a three-vector $\mathbf{k}$ and controlled by a focusing parameter
\begin{equation}
\xi_{\mathrm{foc}} \equiv \frac{\lambda_{\mathrm{dB}}}{R_{\star}} .
\label{eq:foc}
\end{equation}
which is the ratio between the de Broglie wavelength $2 \pi / k$ and $a_{0}$. If a star or BH with mass $M_{BH}$ is embedded in the galactic DM halo with local density $\rho_{dm}$, ultralight DM waves should encounter and scatter the object. For $\xi_{f o c} \ll 1$, the DM behaves like a plane wave and acts as if it is not affected by the external gravitational potential, whereas if $\xi_{f o c} \gg 1$, the DM waves are distorted and amplified close to and behind the astrophysical body. 

The full solution to Eq. \eqref{eq:GP}, including self-interactions, has been computed perturbatively for small $g$ in \cite{budker2023genericformationmechanismultralight}. The DM phase-space distribution is modeled as a Gaussian in terms of the DM virial velocity, $v_{\mathrm{dm}}$ and the local background density $\rho_{\mathrm{dm}}$. These would take different values for different positions in the galactic DM halo. 

By expressing $\psi=\psi^{(0)}+\psi^{(1)}+\psi^{(2)}+\ldots$, where $\psi^{(i>0)}$ indicates the perturbations to $\psi^{(0)}$, and $\psi$ in terms of the sum of the bounded plus the scattering states, $\psi = \psi_{nlm} + \psi_{\textbf{k}}$, the perturbative solution for Eq. \eqref{eq:GA} has been calculated in two steps:
\begin{itemize}
    \item Before any bosons are present in the bound state over times $0 < t < \Delta t$, only terms of order $\psi^{(1)}$ in the perturbative expansion appear in Eq. \eqref{eq:GA} and there is a change in the number of particles in the halo induced by direct capture whose average is given by 
    \begin{equation}
    \label{eq:term1}
    \langle \dot{N}_{n l m} \rangle=\frac{d}{d t}\langle | c_{n l m}^{(1)}(t) |^2 \rangle.
    \end{equation}
    Here $c_{n l m}^{(1)}$ are the first order coefficients in the perturbative expansion, with $l, n, m$ being the quantum numbers for the different hydrogenic states. 
    
    \item For times longer than $t_0 > \Delta t$, a number of bosons are already present in the bound state. These particles can stimulate further capture through Bose enhancement, and be simultaneously depleted via scattering with the background DM waves. In this case, the leading contribution to $\left\langle\dot{N}_{n l m}\right\rangle$ appears at second order such that, 
    \begin{equation}
    \label{eq:term2}
    \begin{split}
    \left\langle \dot{N}_{nlm} \right\rangle
    &= \frac{d}{dt}\left\langle \left| c_{nlm}^{(1)}(t) \right|^2 \right\rangle \\
    &\quad + 2 \sqrt{N_{nlm}^{(0)}} \operatorname{Re} \left\langle \dot{c}_{nlm}^{(2)}(t) \right\rangle.
    \end{split}
    \end{equation}
    \end{itemize}

By adding both Eqs. \eqref{eq:term1} and \eqref{eq:term2}, and multiplying by the mass of the ALP $m_{dm}$, one arrives at the following equation for the rate of change in the halo mass for a particular energy level $(n, l, m)$~\cite{budker2023genericformationmechanismultralight}
\begin{equation}
\label{eq:mass}
    \dot{M}^{n l m}_{GA}=C+\Gamma M^{n l m}_{GA}+\ldots.
\end{equation}
At early times $0 < t < \Delta t$, the mass increases linearly $M^{n l m}_{GA}(t)=C t$, where $C$ represents the mass accreted per unit time. At $t \simeq 1 /|\Gamma|$ the stimulated capture and stripping become relevant. 

If $\Gamma<0$, stripping overcomes stimulated capture and the bound mass saturates to the constant value $M_{\star}^{\text {eq }} \equiv C /|\Gamma|$. If $\Gamma>0$, stimulated capture overcomes stripping, and the mass increases exponentially as $M_{n l m} \propto e^{\Gamma t}$, starting from the initial value $C / \Gamma$. The first case was named as the \textit{dilute GA} case, and the condition for its existence is equivalent to the condition $\xi_{\text {foc }} \equiv 2 \pi \alpha / v_{\mathrm{dm}} \ll 1$. On the other hand, the second case was named as the \textit{dense GA} case, which is equivalent to $\xi_{\text {foc }} \gg 1$.

This way, the spectrum of bound states $\Psi_{nlm}$ and the rate of change $N_{nlm}(t)$ in the number of particles bound to the $nlm$ level were calculated, and the results provide a formation mechanism for the GA that works over timescales longer than the coherence time of the ultralight field $\tau_{dm} \equiv 2\pi/mv^2$. 

For a dense GA forming in the exponential regime, the results presented in \cite{budker2023genericformationmechanismultralight} show that 
\begin{equation}
\label{eq:gamma}
  \begin{aligned}
  \Gamma =  
  &\frac{10 \rho_{\mathrm{dm}}^2}{192 f_a^4 m_{dm}^3 v_{\mathrm{dm}}^2}.
  \end{aligned}
\end{equation}
and the density profile of the halo in its fundamental state is given by $\rho=M_{\mathrm{GA}}^{(1, 0, 0)}(t)\left|\psi^{(1, 0, 0)}\left(r\right)\right|^2$, where 
\begin{equation}
    M_{GA}^{(1, 0, 0)} \approx C_0\exp \left( \Gamma t\right)
\end{equation}
with 
\begin{equation}
\label{eq:c0}
C_0 \equiv \frac{8\pi^{3/2}\rho_{dm}}{m_{dm}^3v_{dm}^3}.
\end{equation}
and $\psi^{(1, 0, 0)}$ is the hydrogen-like wave function in its fundamental state, 
\begin{equation}
    \psi^{(1, 0, 0)} = \frac{1}{\sqrt{\pi a_0^3}}\exp{\left( - \frac{r}{a_0}\right)}
\end{equation}
putting both together, this leads to the profile
\begin{equation}
    \begin{split}
        \rho(r, t)
        &= \frac{C_0}{\pi a^3_0} \exp{\left(\Gamma t\right)}\exp{\left(-\frac{2r}{a_0}\right)}.
    \end{split}
    \label{eq:density_profile}
\end{equation}
In \cite{budker2023genericformationmechanismultralight}, the formalism was applied for an ALP with $f_a = 10^7 \rm GeV$ and $m_{dm} = 10^{-13}\rm eV$, to a star like the Sun, and the results predicted the existence of a GA around the Sun with a Bohr radius of about 1 AU.

Once the density reaches a critical density
\begin{equation}
\label{eq:critrho}
\rho_{\text {crit }} \equiv 16f_a^2m^4_{dm}M_{BH}^2,
\end{equation}
the self-interaction energy is negligible compared to that of gravity (evaluated at $r \simeq R_{\star}$). It has been proven both for the case of Gross-Pitaevskii and in full numerical relativity that regardless of the quartic coupling sign the cloud must saturate \cite{Boudon_2024, Aurrekoetxea_2024}. Therefore, we will assume that a saturated cloud always forms, independently of the sign of the self-interaction coupling.

\begin{figure*}[t]
    \centering
    \begin{subfigure}[t]{0.48\textwidth}
        \centering
        \includegraphics[width=\linewidth]{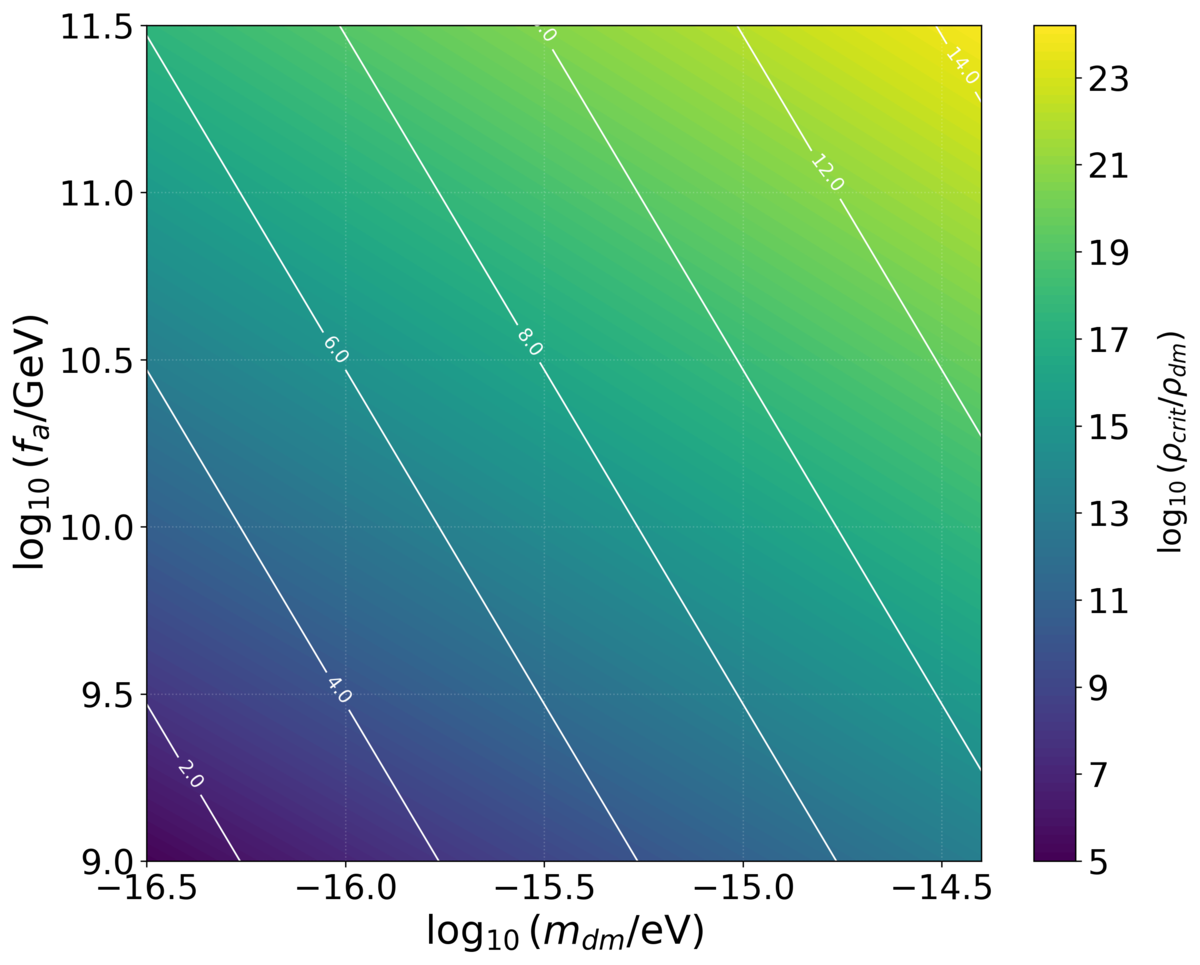}
    \end{subfigure}
    \hfill
    \begin{subfigure}[t]{0.48\textwidth}
        \centering
        \includegraphics[width=\linewidth]{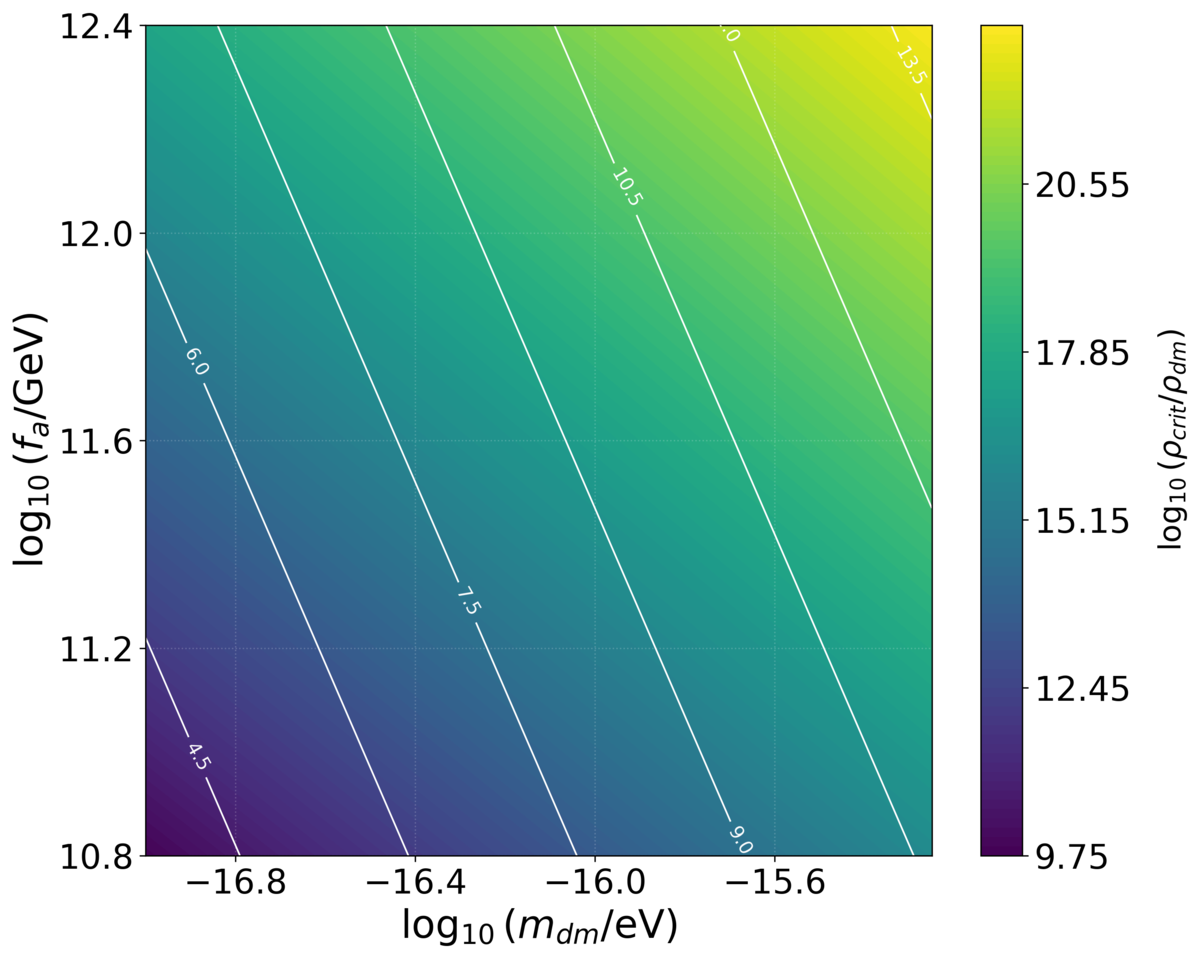}
    \end{subfigure}

    \caption{Contour maps of the critical densities of the clouds for a BH with $M_{BH} = 10^4M_{\odot}$ and $\rho_{dm} = 10^3 GeV/cm^3$ (left) and $M_{BH} = 10^5M_{\odot}$ and $\rho_{dm} = 10^4 GeV/cm^3$(right) as given by \ref{eq:critrho}. The white lines show the different values of the time at which the cloud reaches its critical density $\tau_{crit}$ in Myrs. The critical density increases as $m_{dm}$ and $f_a$ increase. A similar behavior is observed for $\tau_{crit}$, albeit with a slightly rotated gradient direction.}
    \label{fig:dm_densities_times}
\end{figure*}

Fig.~(\ref{fig:dm_densities_times}) shows the contour maps of $\rho_{\rm crit}/\rho_{dm}$, as a function of the DM particle mass $m_{dm}$ and the axion decay constant $f_a$ for two BH masses $M_{BH} = 10^4\,M_\odot$ (left) and $M_{BH} = 10^5\,M_\odot$ (right). In both panels the critical density increases toward the upper-right corner of the parameter space, i.e., for large values of both $m_{dm}$ and $f_a$, consistent with Eq.~\ref{eq:critrho}. Superimposed white contour lines indicate the timescale $\tau_{\rm crit}$ (in Myr) (defined in Eq.\ref{eq:tcrit}) at which the cloud reaches its critical density, which follows a similar but slightly rotated gradient direction with respect to $\rho_{\rm crit}$.  The overall structure of the contour maps is consistent between the two BH masses, with the higher-mass case exhibiting a shifted parameter range toward slightly larger values of $f_a$. Regions where the formation timescale exceeds the age of the universe ($\tau > t_{universe} \simeq 1.38 \times 10^4~\mathrm{Myr}$) are astrophysically irrelevant, as the GA cannot form within a Hubble time.

The GAs described in our work have different characteristics than those generated by superradiance. No rotation is required and, for suitable values of the ALP parameter space, the formation timescales involved are different. For example, in the case where $M_{BH} m_{dm} \ll 1$, the ratio between the generic formation mechanism timescale ($\tau_{\rm GA}\equiv1/\Gamma$) and the superradiant instability timescale ($\tau_{\rm SR}$) is~\cite{Dolan_2007,Brito_2020} 
\begin{equation}
\label{eq:ratio}
\frac{\tau_{\rm GA}}{\tau_{\rm SR}}
\sim
\mathcal{C}\,
\frac{f_a^4\, m_{\rm dm}^{13}\, M_{BH}^9\, v_{\rm dm}^2}{\rho_{\rm dm}^2} \chi\,,
\end{equation}
where $\mathcal{C}$ is a numerical coefficient $\mathcal{O}(\mathcal{C})\sim 1$ and $\chi$ the dimensionless spin of the BH. This relation implies a strong dependence on the BH mass, scaling as $\tau_{\rm GA}/\tau_{\rm SR} \propto M_{BH}^9$. For representative values $\chi \sim 0.5$, 
$f_a \sim 10^{11}\,\mathrm{GeV}$,  $m_{dm} \sim 10^{-16}\, \mathrm{eV}$, in the mass range $M_{BH} \in [10^4,10^8]\,M_\odot$, this ratio spans many orders of magnitude; it is $\ll 1$ for lower masses, indicating that the self-interaction-driven formation mechanism is faster than the superradiant instability, while for sufficiently large masses it becomes $\gg 1$, so that superradiance dominates.

Notice that not only the formation timescales are radically different: the growing mode of the GA generated by self interactions, is in its fundamental state, $n=1$ and $l=0$, whereas the dominant growing mode in the superradiant case has $n=1$ and $l=1$. Thus, they posses different density profiles which can have an impact on their  astrophysical signatures.
\section{SETUP}

In this work we focus on the EMRI scenario, where a GA is forming around a central massive BH, while a smaller companion inspirals within the generated overdensity. 

\begin{figure}[htb]
\centering
\begin{tikzpicture}[scale=1.4,   BHbig/.style={ball color=black},
            BHsmall/.style={ball color=black},
                    DMfill/.style={blue,opacity=#1,draw=none}]
  \def\a{1.0}     
  \def\RDM{1.7}   
  \foreach \i in {1,...,20} {
    \pgfmathsetmacro{\radius}{\i*\RDM*\a/20}
    \pgfmathsetmacro{\opacity}{0.9*exp(-2*\radius/\a)}
    \fill[DMfill={\opacity}] (0,0) circle (\radius);
  }
  \shade[BHbig] (0,0) circle (0.30);
  \node[below=2pt] at (0,-0.25) {\small $M_{BH}$};
  \draw[->, red]
  node[pos=1, scale=0.75, above=0.01pt, right=22pt, red] {$a_0$};  
  \pgfmathsetmacro{\radio}{0.8*\a};
  \draw[red, dashed] (0,0) circle (\radio);
  \coordinate (mTwo) at (1.25*\a,0.55*\a);
  \shade[BHsmall] (mTwo) circle (0.12);
  \node[above right=2pt] at (mTwo) {\small $m$};
  \pgfmathsetmacro{\rSmall}{veclen(1.25*\a,0.55*\a)}
  \draw[black, dashed] (0,0) circle (\rSmall);
  \draw[->, thick, black, opacity=0.45]
    (0,0) -- (1.25*\a, 0.55*\a)
    node[midway, above=3pt, opacity=1] {\small $r_2$};
\end{tikzpicture}
\caption{An EMRI with a central BH of mass $M_{BH}$ creating a bosonic cloud overdensity and a secondary BH of mass $m$ is placed with its approximate circular orbit shown as a dashed circle. We expect that $r_2 \gtrsim a_0$ in order to see the environmental imprint in the binary evolution.}
\label{fig:setup}
\end{figure}
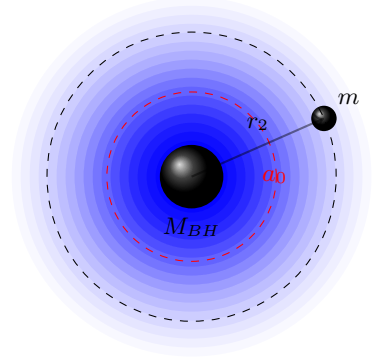

Fig.~\ref{fig:setup} illustrates an EMRI with a primary BH mass $M_{BH}$ and a secondary with mass $m$. Around the massive BH a GA is forming from  a local background DM density $\rho_{dm}$, where $a_0$ is the characteristic cloud radius. The secondary BH orbits the primary on an approximately circular orbit, depicted as a dashed circle, at a separation $r_2 \gtrsim a_0$. As it moves through the cloud, it experiences environmental effects that alter its orbital evolution, providing a potential observational signature of the GA through GW emission.

The DM overdensity seen by the secondary BH is going to be equal to the local density of the GA formed in the central supermassive BH, evaluated at the companion's position plus the background DM density, which will be negligible with respect to the former for suitable regions of parameter space. Therefore, from Eq. \eqref{eq:density_profile} and using Kepler`s law $r = r_2 =  \left(M_{BH}/(\pi f_{\rm GW})^2\right)^{1/3}$, we obtain
\begin{multline}
\label{eq:density_profile_f}
\rho(f_{GW}, t) = \rho_{GA} + \rho_{dm} = \\
\frac{C_0}{\pi a^3_0} \exp{\left(\Gamma t\right)}\exp{\left(-\frac{2}{a_0}\frac{M_{BH}^{1/3}}{(\pi f_{GW})^{2/3}}\right)} + \rho_{dm}.
\end{multline}
Next, we estimate the timescale over which the GA reaches saturation density. This critical time $\tau_{crit}$ can be obtained by equating the GA density at $r=0$ to the critical density at such time, namely  $\rho(r=0, \tau_{crit})=\rho_{crit}$, which yields  
\begin{equation}
\label{eq:tcrit}
\tau_{crit}=\frac{64 m_{dm}^3 f_a^4 v_{\mathrm{dm}}^2}{\rho_{\mathrm{dm}}^2} \ln \left(\frac{2 \pi^{-1 / 2} m_{dm} f_a^2 v_{d m}^3}{M_{BH} \rho_{d m}}+1\right).
\end{equation}
The orbital evolution of the binary can be calculated from an energy balance equation. Since the dissipation timescale is much longer than the orbital timescale, the system evolves adiabatically, and the orbital evolution is determined by the total power loss from both GW emission and environmental effects,
\begin{equation}
    \dot E = -\dot{E}_{\rm GW}\,-\dot{E}_{\rm DF}-\dot{E}_{\rm ACC},
    \label{eq:energy_conservation}
\end{equation}
where $\dot{E}_{\rm GW}$ is the energy loss caused by GW emission, $\dot{E}_{\rm DF}$ is the energy loss caused by dynamical friction (DF) and $\dot{E}_{\rm ACC}$ is the energy loss caused by (ACC). Here $E$ is the total orbital energy, given by 
\begin{equation}
\label{eq:Eorb}
    E = -\frac{1}{2} \mathcal{M}_c^{5/3} M^{-2/3} v^2
\end{equation}
and the power emitted by GWs for a circular binary at leading Post-Newtonian order is given by the standard quadrupole formula \cite{Maggiore2008}, namely
\begin{equation}
\label{eq:Egw}
    \dot{E}_{\rm GW} = \frac{32}{5}\,\mathcal{M}_c^{10/3}\,M^{-10/3}\,v^{10}
\end{equation}
For a realistic accretion process \footnote{We have also considered the generic formation mechanism in Eq.~\eqref{eq:mass} as an ACC effect, leading to the perturbative solution  $    \dot{m} = C_0 + \Gamma m$. However, we have verified that the accretion described by this formation process is completely negligible, since the binary evolution timescale is not sufficient to generate a significant GA overdensity around $M_{BH}$. }, arising from flows of particles falling into the potential of the companion BH, standard benchmarks such as the Bondi-Hoyle-Lyttleton accretion rate \cite{Bondi:1952ni, Bondi:1944rnk} can be approximately applied for self-interacting ultralight bosons \cite{Boudon_2024}. This accreted mass rate can then be written in terms of the density in Eq. \eqref{eq:density_profile} evaluated at the companion position $r_2$, namely
\begin{equation}
\label{eq:dotMass}
\dot{m}=\frac{4 \pi \lambda m^2 \rho(r_2(f), t)}{(v^2 + c_s^2)^{3/2}}.
\end{equation}
Here $\lambda$ is a dimensionless fudge factor of order unity that accounts for the detailed hydrodynamics of the ACC flow, which we took to be $\lambda = 1$, and $c_s$ is the sound velocity of the halo \cite{Boudon_2024}, which is given by 
\begin{equation}
\label{eq:sound}
    c_s^2 =3 \lambda  \frac{\rho_{GA}(r_2, t)}{4 m_{dm}^4}.
\end{equation}
We have checked that, for all the explored parameter space $(f_a, m_{dm})$ where we have numerically found dephasing, $v/c_s \gtrsim \mathcal{O}(10)$ and $(v^2 + c_s^2)^{3/2} \approx v^3$, at least for the formation timescales considered in this work.

In this context, Eq. \eqref{eq:dotMass} can be used to compute the power emitted by ACC, assuming that $m$ in the total orbital energy depends on time, that is, 
\begin{equation}
\begin{aligned}
\label{eq:ACCdot}
    \dot{E}_{\rm ACC} = \frac{1}{2}\,\frac{dm}{dt}\,v^{2}
\end{aligned}
\end{equation}
DF is caused by the gravitational pull of the smaller BH, which induces a DM overdensity in its wake. This overdensity exerts an attractive force that opposes the motion, causing the BH to lose energy and momentum. This effect, first studied by Chandrasekhar for a medium composed of collisionless particles with non-relativistic velocities \cite{Chandra_DF},  was found to be 
\begin{equation}
\label{eq:Edf}
\dot{E}_{\mathrm{DF}}=\frac{4 \pi m^2 \rho(r_2, t)}v \mathcal{I}[f_{\text{GW}}] .
\end{equation}
Here $\mathcal{I}$ is the Coulomb factor which, for ultralight bosons, takes the following form~\cite{Hui_2017,Traykova_2023}
\begin{equation}
    \mathcal{I}(x) = \mathrm{Cin}(x) + \mathrm{sinc}(x) - 1,
    \qquad
    x = 2\,m_{\rm dm}\,r\,v,
\end{equation}
where $\mathrm{sinc}(x) = \sin(x)/x$ and $\mathrm{Cin}(x)$
\begin{equation}
    \begin{split}
    \mathrm{Cin}(x) &= -\int_x^{\infty} \frac{\cos t}{t}\,\mathrm{d}t \\
    &= \gamma_{\mathrm{E}} + \ln x + \int_0^x \frac{\cos t - 1}{t}\,\mathrm{d}t,
    \end{split}
\end{equation}
with $\gamma_{\rm E}$  the Euler--Mascheroni constant.
By using Eq. \eqref{eq:energy_conservation} and the different expression for the energy losses previously discussed \eqref{eq:Egw}, \eqref{eq:Edf} and \eqref{eq:ACCdot}, we obtain a differential equation for the evolution of the GW frequency \cite{CanevaSantoro:2023aol}, namely
\begin{equation}
\label{eq:ODE}
    \begin{split}
        \dot{f}_{\text{GW}} &= \frac{96}{5} \pi^{8/3} \mathcal{M}_c^{5/3} f_{\mathrm{GW}}^{11/3}\\
        &\quad + 12\, q \,\rho(f_{GW}, t) |\mathcal{I}[f_{\text{GW}}]| + 6\, q \,\rho(f_{GW},t)
    \end{split}
\end{equation}
This equation can be solved numerically to obtain the GW frequency. The phase can be calculated integrating in time 
\begin{equation}
\label{eq:phase}
\phi(t)=2 \pi\int_{t_{\min }}^{t_{\max }} f_{\mathrm{GW}}(t) d t,
\end{equation}
and the number of cycles would be $\mathcal{N}_{\mathrm{cyc}}= \phi(t)/2 \pi$. The environmental contribution effectively enters the phase evolution with a negative power of the orbital velocity, corresponding to a $-5.5\text{PN}$ negative PN correction (see Appendix~\ref{sec:appendix} for details). Importantly, this contribution has a frequency dependence different from the leading vacuum term, $\phi_{\rm vac}(f)\propto \mathcal{M}_c^{-5/3} f_{GW}^{-5/3}$, and therefore cannot be absorbed into a simple redefinition of the chirp mass $\mathcal{M}_c$.

Therefore, the impact of the halo on the inspiral is determined by the interplay between the growth timescale of the GA, the coalescence timescale of the binary, and the critical timescale associated with the saturation of the halo. Observable effects arise when the halo reaches sufficiently large densities many cycles before merger. We explore the parameter space spanned by $f_a$, $m_{dm}$ and $M_{BH}$, fixing $\rho_{dm}$ to realistic values and $m = 10M_{\odot}$. In our work we have fixed $v_{dm} = 2.4\times 10^5 m/s$, noting that these velocities should change at different galactic positions, but that choosing different values within a realistic range does not significantly affect our results. 

\subsection{PARAMETER RANGES}
\label{sec:par_ranges}

The range of possible $m_{dm}$ will be determined by the size of the central BH and the coalescence time. In principle, there are no strict lower bounds for the ALPs masses coming from particle physics; however, the de Broglie wavelength should not exceed the size of the smallest observed structures, giving $m_{dm} \gtrsim 10^{-22}$ eV \cite{Ferreira_2021}, while the Lyman-$\alpha$ forest pushes this bound up to $m_{dm} \gtrsim 10^{-21}$ eV \cite{Ir_i__2017}.  To find observable effects, we need at least $a_0\geq R_s$, where $R_s$ is the Schwarzschild radius of the BH. This sets a minimum $m_{dm}$ which we will be able to constrain
\begin{equation}
\label{eq:mdm_min}
m_{\rm dm}^{min} = \frac{1}{\sqrt{2}M_{BH}}.
\end{equation}
The minimum values, for the central BH masses explored, are
\begin{center}
\begin{tabular}{cc}
\hline
$M_{\rm BH}$ & $m_{dm}^{min}$ \\
\hline
$10^4\,M_\odot$ & $1.1 \times 10^{-14}$\,eV \\
$10^5\,M_\odot$ & $3.47 \times 10^{-14}$\,eV \\
$10^6\,M_\odot$ & $1.1 \times 10^{-15}$\,eV \\
$10^7\,M_\odot$ & $3.47 \times 10^{-15}$\,eV \\
\hline
\end{tabular}
\end{center}
More precisely, in order to work with a GA description based on a Newtonian approximation, we should expect $a_0 \gg R_s$. However, for all of the massive BH populations explored in this work, this condition is naturally satisfied within the $m_{\mathrm{dm}}$ range where measurable dephasing occurs, mainly because the mass range capable of generating a halo overdensity within a realistic timescale is primarily determined by the condition that the de Broglie wavelength ($\lambda_{dB}\sim 1/(m_{dm} v_{dm})$) is of the same order as the BH gravitational radius.

\begin{figure}    \includegraphics[width=1\linewidth]{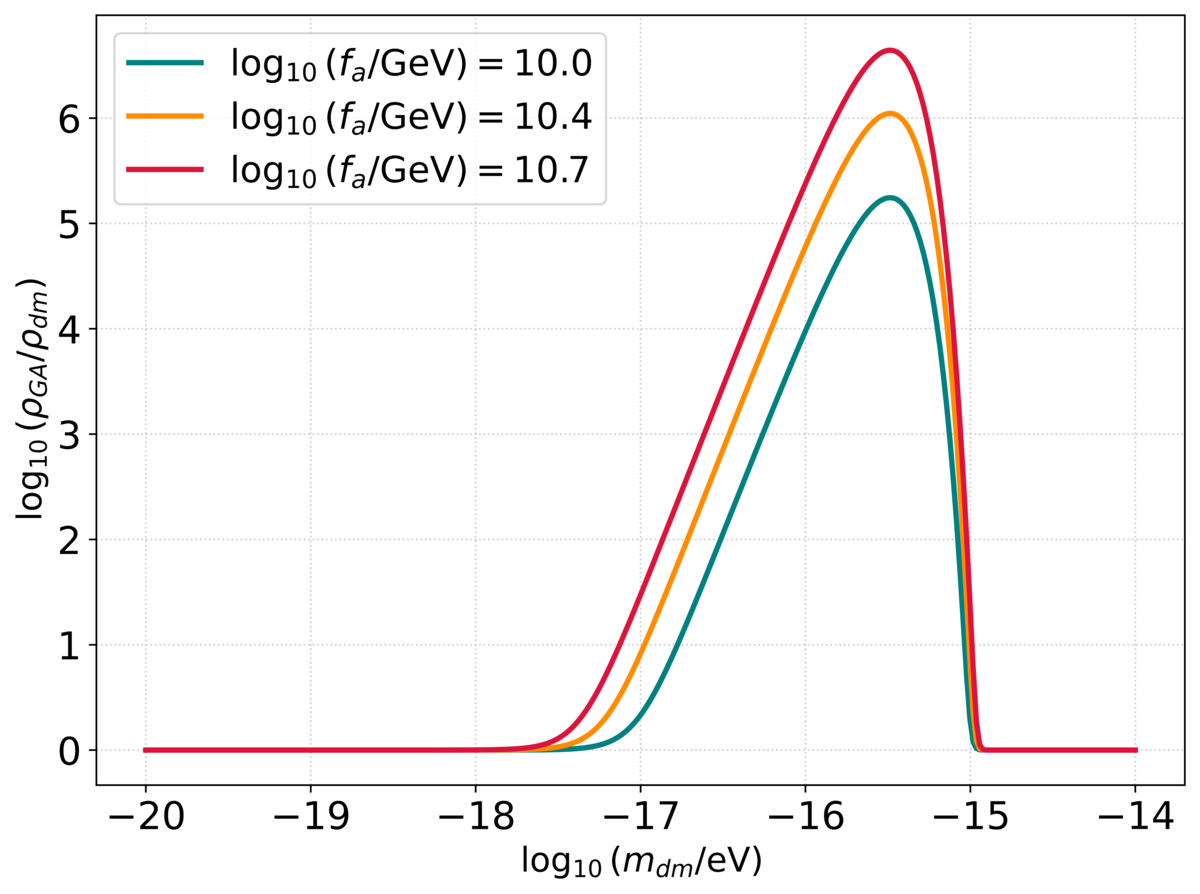}
    \caption{Density profile ratios taken at the initial orbital position $r(f_0 = 10^{-4} \text{Hz})$ and a BH mass of $M_{BH} = 10^5M_{\odot}$ as a function of the particle $m_{\mathrm{dm}}$. The overdensity appears within specific ranges of $m_{\mathrm{dm}}$, consistent with the physical limits of the profile: at low $m_{\mathrm{dm}}$, the $1/a_0^3$ factor in Eq. \eqref{eq:density_profile} suppresses the overdensity since $a_0 \propto m_{\mathrm{dm}}^{-2}$, whereas at high $m_{\mathrm{dm}}$, the suppression is driven by the $e^{-2r/a_0}$ term. Combined with the GA formation timescales set by $f_{\mathrm{a}}$. } 
    \label{fig:rho_mdm}
\end{figure}

In Fig.\ref{fig:rho_mdm} we show the ratio $\rho_{\rm GA}/\rho_{dm}$ of the GA density, as given by Eq.(\ref{eq:density_profile}), to the background DM density, for a BH with $M_{BH} = 10^{4}M_{\odot}$ and background density $\rho_{dm} = 10^4\text{GeV/cm}^3$. This ratio, displayed as a function of the DM particle mass $m_{\rm dm}$,  is  evaluated at a orbital separation such that $f(r_2) = 10^{-4}\text{Hz}$  for three representative values of the ALP decay constant $f_a$. The overdensity is confined to a specific window in $m_{\rm dm}$. This is because, at low $m_{\rm dm}$, $\rho_{GA}(f(r_2), t)$ is suppressed by the $1/a_0^3$ factor, since $a_0 \propto m_{\rm dm}^{-2}$ causes the cloud to become increasingly diffuse. Conversely, at high $m_{\rm dm}$, the exponential factor $e^{-2r/a_0}$ drives a rapid suppression as the cloud becomes more compact than the orbital separation. The resulting peaked structure shifts toward lower $m_{\rm dm}$ and reaches different amplitudes for different values of $f_a$, consistent with changes in the GA formation timescales imposed by this parameter. This behavior directly underlies the triangular morphology seen in the parameter space maps of Fig.~\ref{fig:dm_contours_phi}.
On the other hand, the decay constant $f_a$ is a free parameter for ALPs, with a reasonable range being
\begin{equation}
    10^{8} \text{ GeV} \lesssim f_a \lesssim 10^{18} \text{ GeV},
\end{equation}
where the lower bound comes from astrophysical constraints such as stellar cooling and SN1987A by assuming a coupling to matter \cite{Raffelt}. We do not consider any extra couplings here and will not entertain such small values for this parameter. The upper bound comes from the requirement that the effective field theory remains valid below the Planck scale \cite{Marsh_2016}. In practice, the main impact that the strength of the self interactions, mediated by $\lambda = (m_{dm}/f_a)^2$, have in our results, is in how they affect the formation timescale, via $\Gamma$ and $\tau_{crit}$.

The broad range of possible values for $f_a$ and $m_{dm}$ allows us to find very different values for $\rho_{crit}$ and thus different $\tau_{crit}$. In the most general case, the GA formation starts at a different time than the start of the binary inspiral. We defined this timescale as $t_0$.

IMBHs or EMBHs, with $10^4 \lesssim M_{\rm BH}/M_\odot \lesssim 10^7$ are expected to arise at $z \gtrsim 10$ through several channels, including remnants of massive Population-III stars forming in minihalos \cite{Madau_2001} \cite{Heger_2003}, runaway stellar mergers in dense primordial clusters \cite{Portegies_Zwart_2002}, and direct-collapse scenarios in atomic cooling halos \cite{Loeb_1994}. In standard cosmology, objects formed at $z \sim 10$--$30$ have lookback times of $\sim 13.3$--$13.7$~Gyr, corresponding to $\sim (0.96$--$0.99)\,t_{universe}$. We thus have chosen a reasonable value for the GA formation starting time of $t_0 \sim 0.9 t_{universe}$. Our results are not significantly affected by a choice of, e.g., $t_0 \sim 0.96 \, t_{universe}$, or less than $0.9 \, t_{\mathrm{universe}}$; however, the timescales involved must all be of cosmological order in order to produce sufficiently dense GAs.

We focus on high but realistic $\rho_{dm}$ values. In the inner regions of the Galaxy, the DM distribution is expected to follow a centrally concentrated profile, such as a cuspy Navarro-Frenk-White profile \cite{Navarro_1997} 
\begin{equation}
\label{eq:NFW}
\rho_{\rm NFW}(r) =
\frac{\rho_s}{(r/r_s)\left(1+r/r_s\right)^2},
\end{equation}
where $\rho_s$ is a characteristic density and $r_s$ is the scale radius of the halo. Typical DM densities therefore increase rapidly toward the galactic center, reaching values orders of magnitude larger than in the outer regions of the galaxy. For a Milky-Way-like galaxy, typical values are $r_s \sim 5\;\mathrm{kpc}$, with a normalization mass $M_0 = (4/3\pi r_s^3)\rho_s \sim 3 \times 10^{11} M_{\odot}$ and thus $\rho_s \sim 20\;\mathrm{GeV\,cm^{-3}}$ \cite{ou2023darkmatterprofilemilky}.
At radii much smaller than the scale radius ($r \ll r_s$), the profile approaches a cuspy behavior,
\begin{equation}
\label{eq:NFW_estimation}
\rho_{\rm NFW}(r) \simeq \rho_s \frac{r_s}{r}.
\end{equation}
Evaluating this expression at, for example, $r \sim 1\;\mathrm{pc}$ yields
\begin{equation}
\label{eq:NFW_estimation}
\rho_{\rm NFW}(1\,\mathrm{pc})
\sim \rho_s \frac{5\,\mathrm{kpc}}{1\,\mathrm{pc}}
\sim 10^4\text{-}10^5\;\mathrm{GeV\,cm^{-3}},
\end{equation}
In principle, we can achieve larger values if we go to smaller values of $r$. As we explore different values of $M_{BH}$, we choose the smallest order-of-magnitude $\rho_{dm}$ that produces a detectable effect over a large region of parameter space. We can increase the reachable GA densities considered in this work if we assume that higher values can be achieved by the NFW model, or if we assume that the IMRI or EMRI we are focusing on is located near the vicinity of a supermassive BH in a galactic center, where we expect to find a DM overdensity not described by the NFW halo model itself, but caused by, for example, the vicinity of a DM spike. We note however that the formation formalism will produce a detectable GA as long as $\rho_{dm} \ll \rho_{crit}$. As we explain in section \ref{sec:Detectability and distinguishability}, we have found that, for bigger BHs, we needed to increase $\rho_{dm}$ in order to obtain an observable effect. 

\subsection{DEPHASING}

We have solved numerically Eq. \eqref{eq:ODE} and obtained the phase by integrating the numerical solution as in Eq. \eqref{eq:phase}. We used an adaptive step-size integrator based on the LSODA Python algorithm, which automatically switches between stiff and non-stiff methods, with relative and absolute tolerances set to $10^{-10}$ and $10^{-15}$, respectively. The integration was performed over an interval such that the initial frequency is $f_0 = 10^{-4}Hz$ and the final frequency is the one at the Innermost Stable Orbit (ISCO), $f_{ISCO}$. For reference, for a BH of mass $M_{BH} = 10^{4}M_{\odot}$ is $f_{ISCO} \simeq 0.43Hz$. 

\begin{figure*}[t]
    \centering
    \begin{subfigure}{0.48\textwidth}
        \centering
\includegraphics[width=\linewidth]{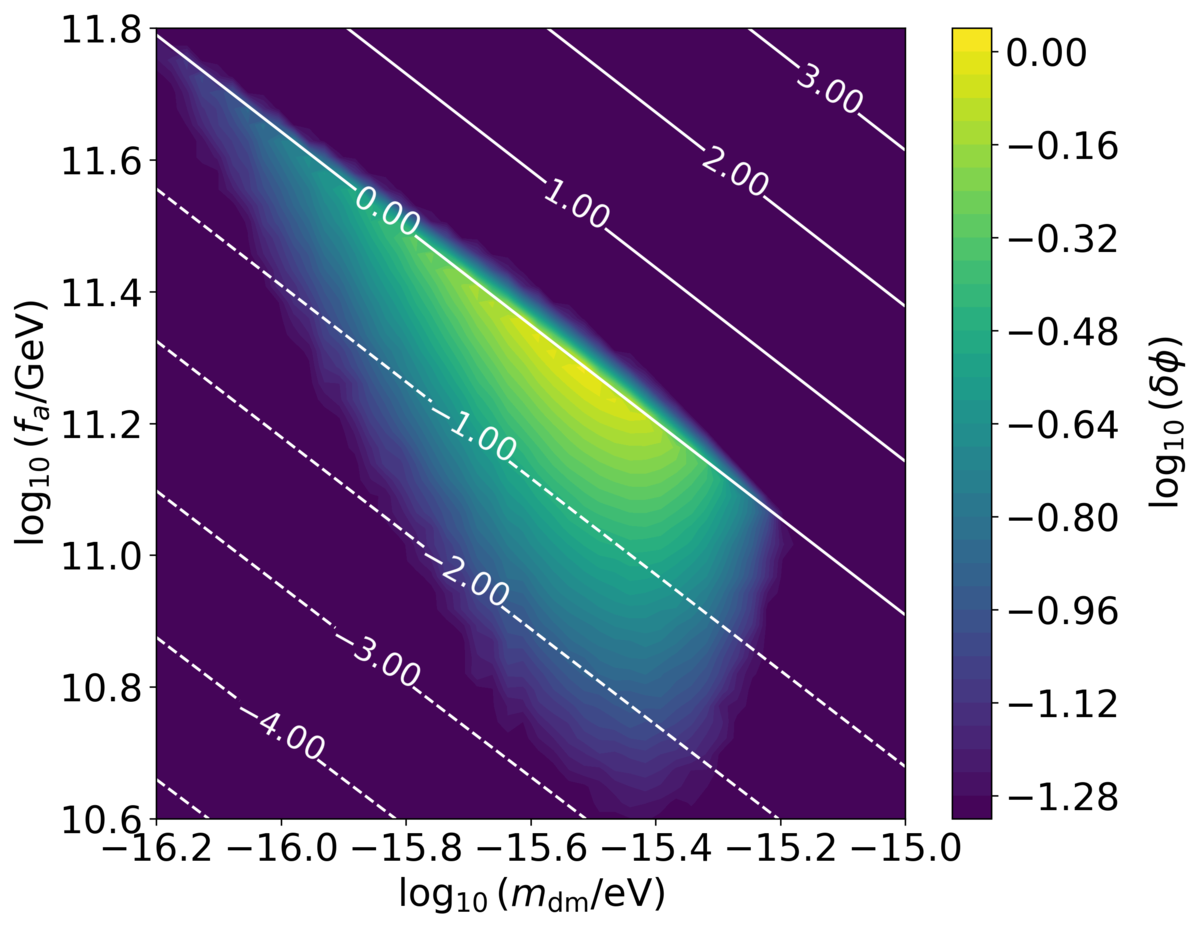}
    \end{subfigure}
    \hfill
    \begin{subfigure}{0.48\textwidth}
        \centering        \includegraphics[width=\linewidth]{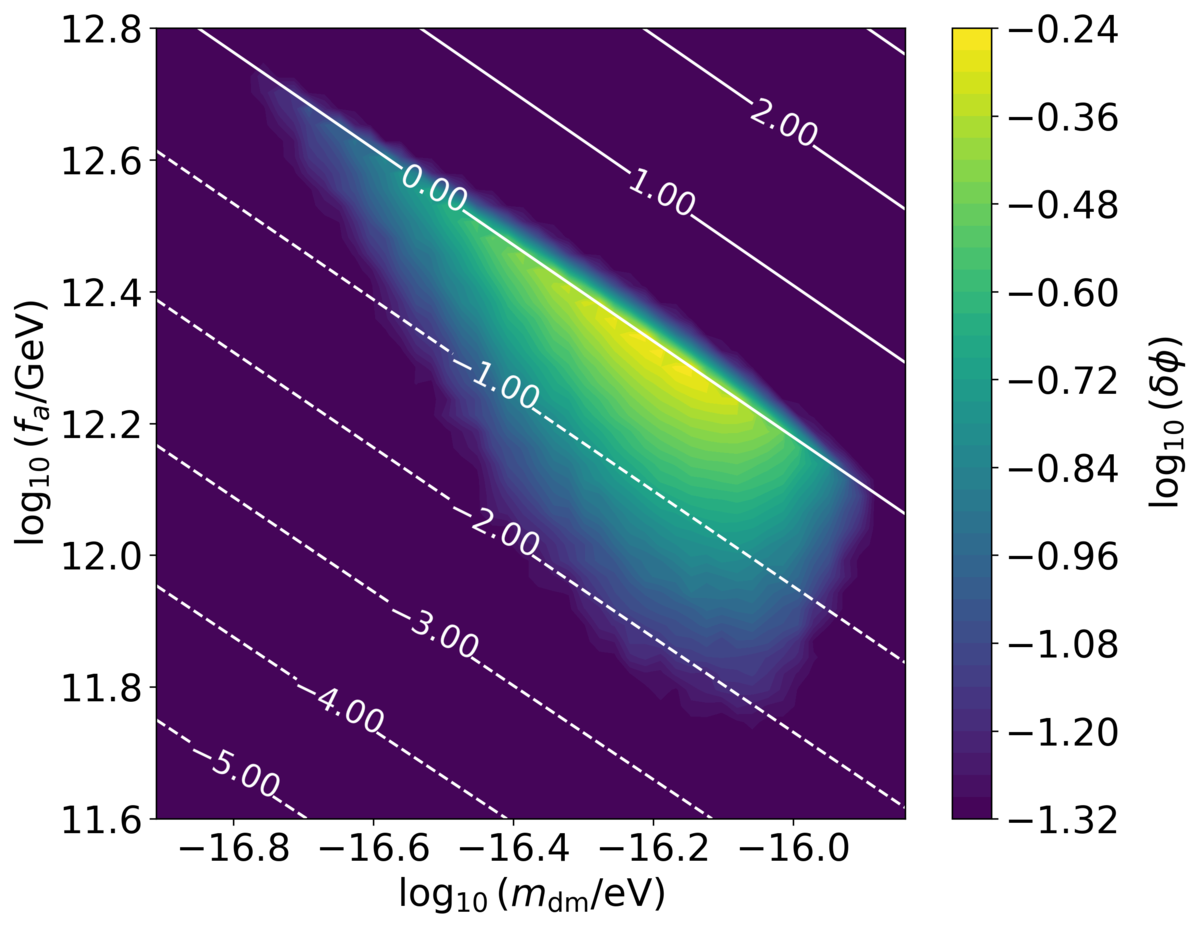}
    \end{subfigure}
    \caption{Maximum accumulated dephasing, defined as $\max(\phi_{DF}-\phi_{B})$ with $\phi_{DF}$ and $\phi_{V}$ being the phase of the BBH immersed in the GA cloud and in the background $\rho_{dm} = 10^4 GeV/cm^3$ and $\rho_{dm} = 10^5 GeV/cm^3$, for $M_{BH} = 10^4 M_{\odot}$ (left) and $M_{BH} = 10^5 M_{\odot}$ (right) respectively. White lines denote the quantity $\frac{\tau_{crit}}{t_{0}}$. The region where we find visible dephasing forms a triangular shape. This shape is explained both by the dependence of $\rho_{GA}$ in $m_{dm}$, and the formation timescales involved for the whole combination of parameters.} 
    \label{fig:dm_contours_phi}
\end{figure*}

Fig.~\ref{fig:dm_contours_phi} shows the maximum dephasing computed as $ \delta \phi =\text{max}(\phi_{DF} - \phi_{B})$ between binaries embedded in the GA overdensity and in the background DM density $\rho_{dm}$ for different values of $f_a$ and $m_{dm}$. The integration is performed over a ten years window, chosen between $f_0$ and $f_{ISCO}$, such that the maximum dephasing is achieved. The left panel is for $M_{BH} = 10^4 M_{\odot}$ and the right panel for $M_{BH} = 10^5 M_{\odot}$. For the first panel the maximum dephasing is obtained at values around $m_{dm} = 3.2 \cdot 10^{-15}$ eV and $f_a = 1.58 \cdot 10^{11}$ GeV, whereas for the second panel, this happens at $m_{dm} = 6.3 \cdot 10^{-17}$ eV and $f_a = 2 \cdot 10^{12}$ GeV. The white contour lines denote the regions with different values of $\tau_{crit}/t_{0}$. Reducing $t_0$ several orders of magnitude lowers the white line at $\tau_{crit}/t_0 = 1$, shrinks the region in parameter space with larger dephasing (coloured region) and lowers the maximum dephasing value. The maximum dephasings involved shown in Fig.~\ref{fig:dm_contours_phi} (left), are of the order of les than one orbital cycle, whereas the maximum density is of order the critical density $\rho_{crit} =7.25 \cdot 10^{11}GeV/cm^3$. Several cycles of dephasing can be obtained for these and other central BH masses if we increase $\rho_{dm}$, always keeping it lower than the achieved GA density. The achieved dephasing is detectable by LISA, as we show in section \ref{sec:Detectability and distinguishability}.

Finally, the triangular shape with a diagonal cut in the white line arises both because the involved formation timescales, as it is shown in Fig.~\ref{fig:dm_densities_times} as given by  Eq. \eqref{eq:tcrit}, and because of the dependence of the density profile on the particle parameters given by equation \eqref{eq:density_profile} and ilustrated in Fig. \ref{fig:rho_mdm}. 
\section{Detectability and distinguishability}\label{sec:Detectability and distinguishability}
In this section we focus on two distinct but related questions: first, whether LISA is sensitive enough to detect these signals at all; and second, whether the DM-induced corrections to the waveform are distinguishable, both between waveforms generated for different pairs of $(m_{dm}, f_a)$, or from those of a background DM inspiral, which for small realistic values corresponds effectively to a vacuum inspiral. A figure of merit of GW analysis is the mismatch between any two waveforms with strains $h_1$ and $h_2$,
\begin{equation}
    \mathcal{M} \equiv 1- \frac{\left\langle h_1 \mid h_2\right\rangle}{\sqrt{\left\langle h_1 \mid h_1\right\rangle\left\langle h_2 \mid h_2\right\rangle}}.
\end{equation}
The noise-weighted inner product is given by
\begin{equation}
    \left\langle h_1 \mid h_2\right\rangle=4 \operatorname{Re} \int_{f \min }^{f \max } \frac{\tilde{h}_1(f) \tilde{h}_2(f)}{S_n(f)} \mathrm{d} f,
\end{equation}
with $\tilde{a}(f)$ denoting the Fourier transform of $a(t)$ and $S_n(f)$ the one-sided LISA noise power spectral density~\cite{amaroseoane2017laserinterferometerspaceantenna}. We define the SNR of a GW signal $h(t)$ as
\begin{equation}
    \text{SNR}^2 \equiv\langle h \mid h\rangle .
\end{equation}
The distinguishability  of two waveforms is defined as (see Appendix \ref{sec:appendix} for a derivation),
\begin{equation}
\label{eq:detectability_condition}
\text{SNR} \gtrsim \frac{D}{\sqrt{2\mathcal{M}}}\,,
\end{equation}
where $D$ is the dimension of the parameter space considered. This quantity has a direct physical interpretation as the minimum SNR required to distinguish between two waveforms at $1-\sigma$ confidence level. In our case, these are waveforms from systems immersed in the GA overdensity, $h_{GA}$, and waveforms from systems immersed in the local DM background, $h_{BG}$. As we did for the dephasing, we have computed the mismatch $\mathcal{M}(h_{BG}|h_{GA})$, by aligning the signals during 10 years of their respective evolution, choosing the time window so as to constrain the largest region of parameter space. 

\begin{figure*}[t]
    \centering
    \begin{subfigure}[t]{0.48\textwidth}
        \centering        \includegraphics[width=\linewidth]{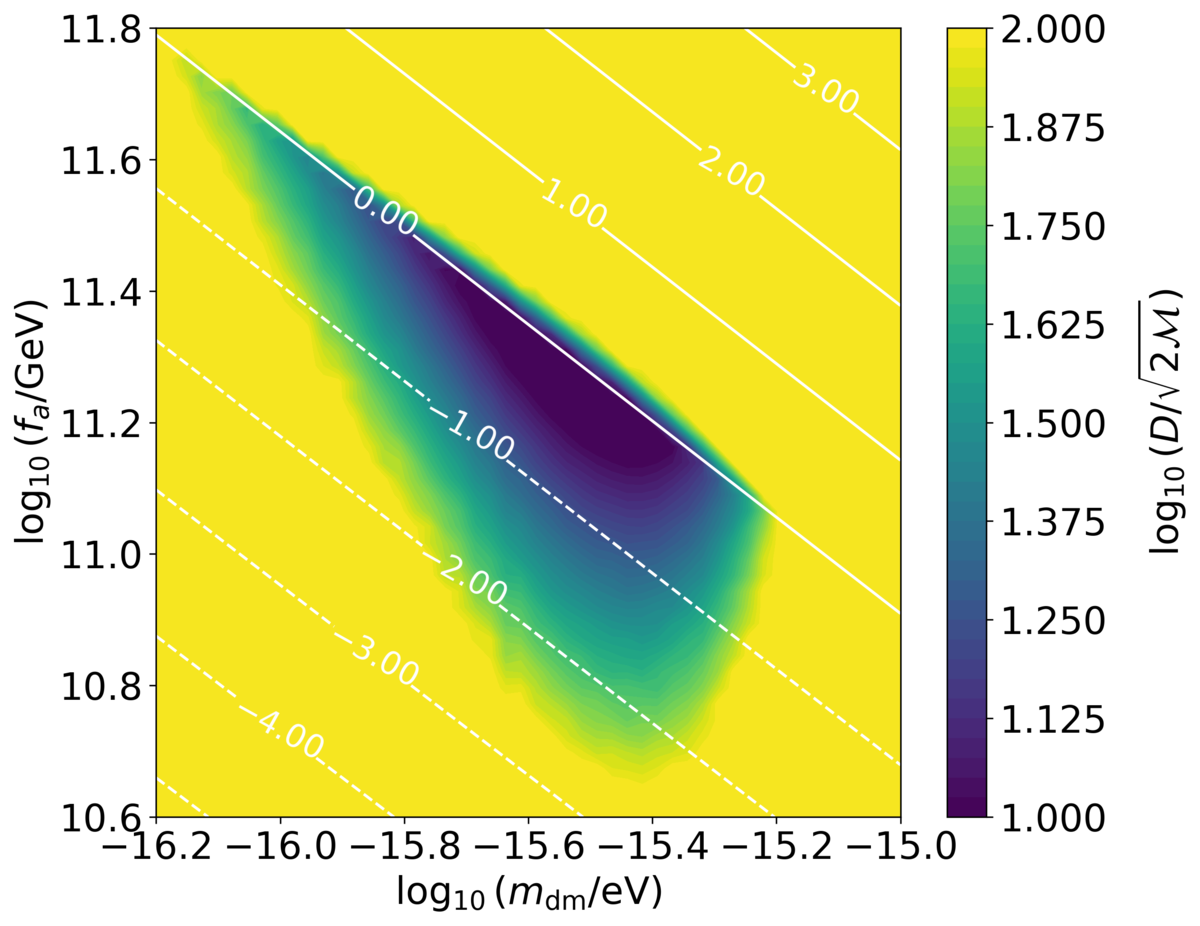}
    \end{subfigure}
    \hfill
    \begin{subfigure}[t]{0.48\textwidth}
        \centering        \includegraphics[width=\linewidth]{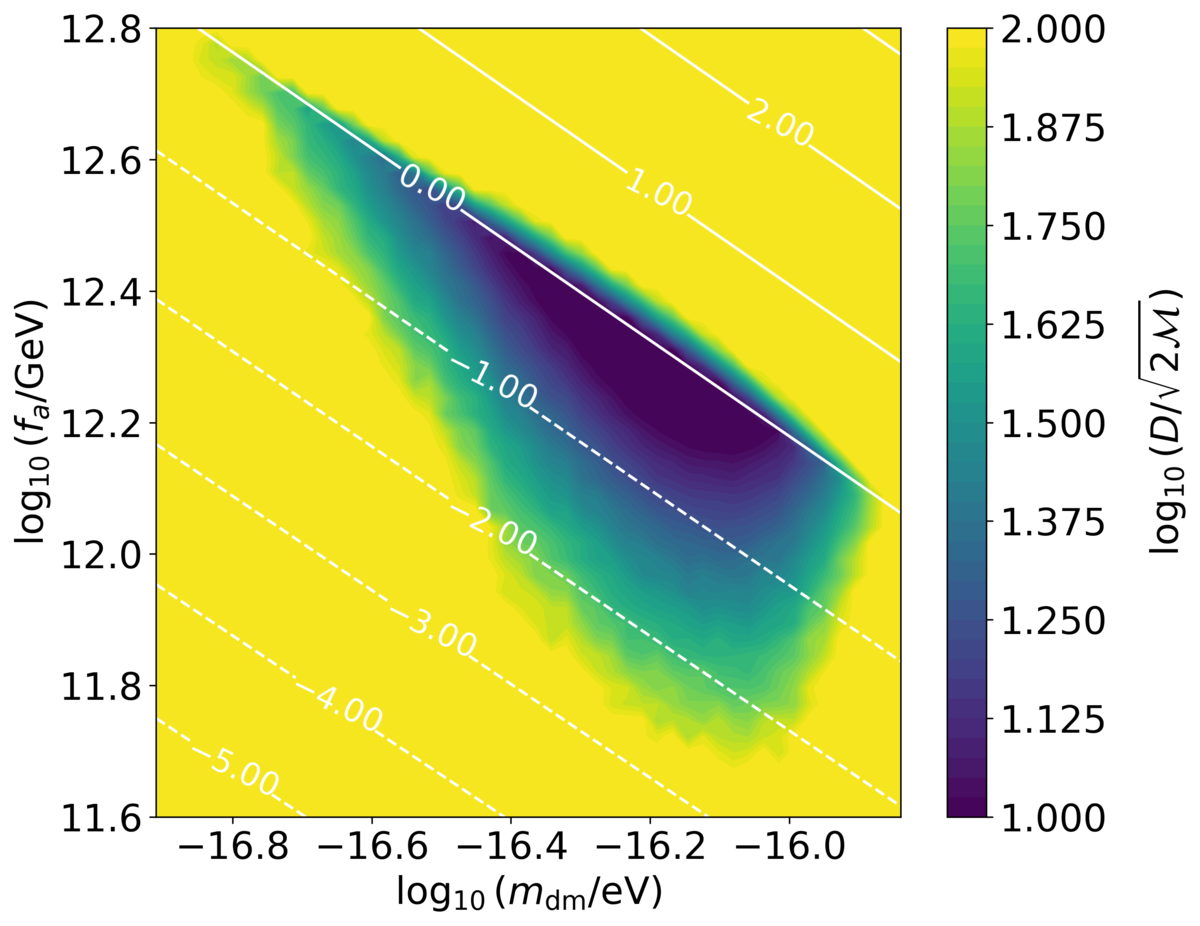}
    \end{subfigure}
    \begin{subfigure}[t]{0.48\textwidth}
        \centering        \includegraphics[width=\linewidth]{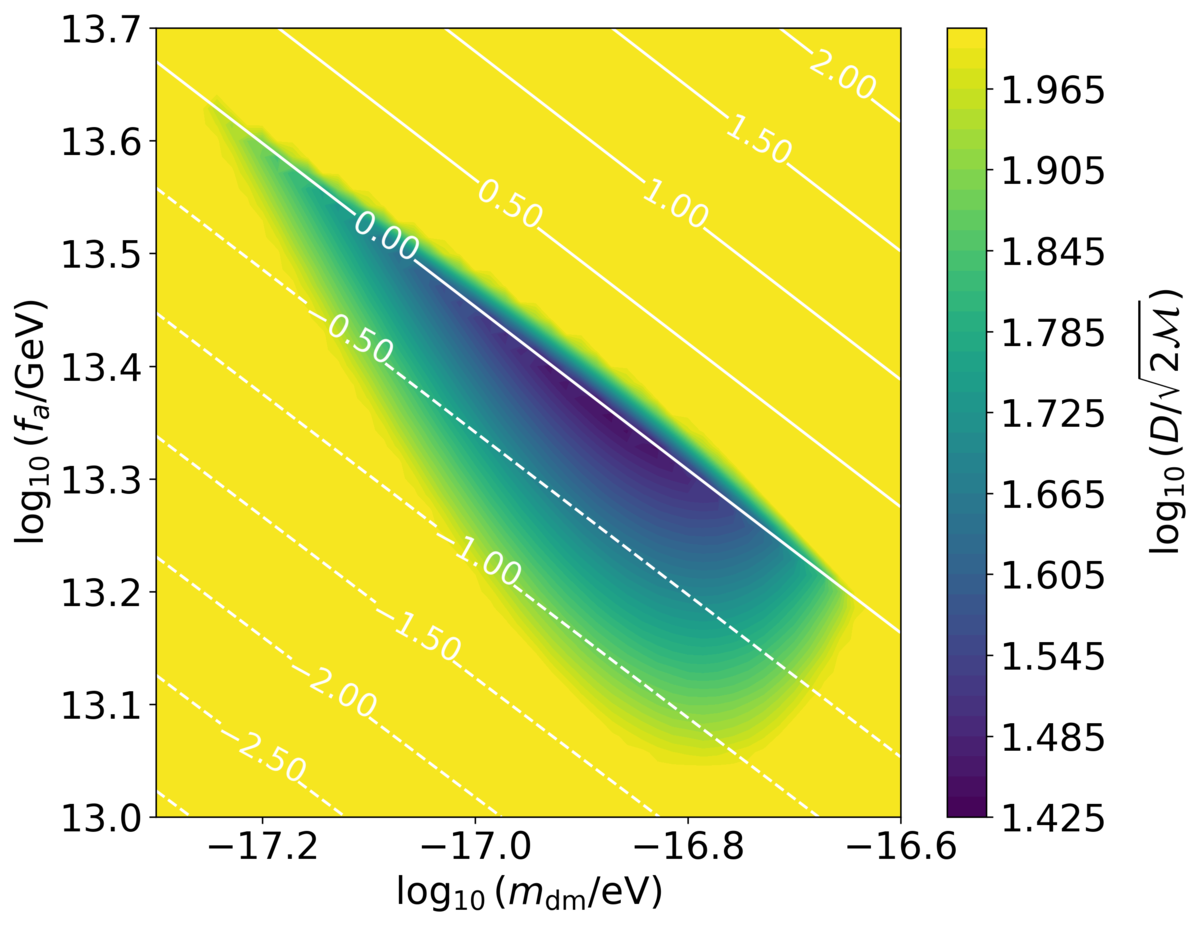}
    \end{subfigure}
    \hfill
    \begin{subfigure}[t]{0.48\textwidth}
        \centering       \includegraphics[width=\linewidth]{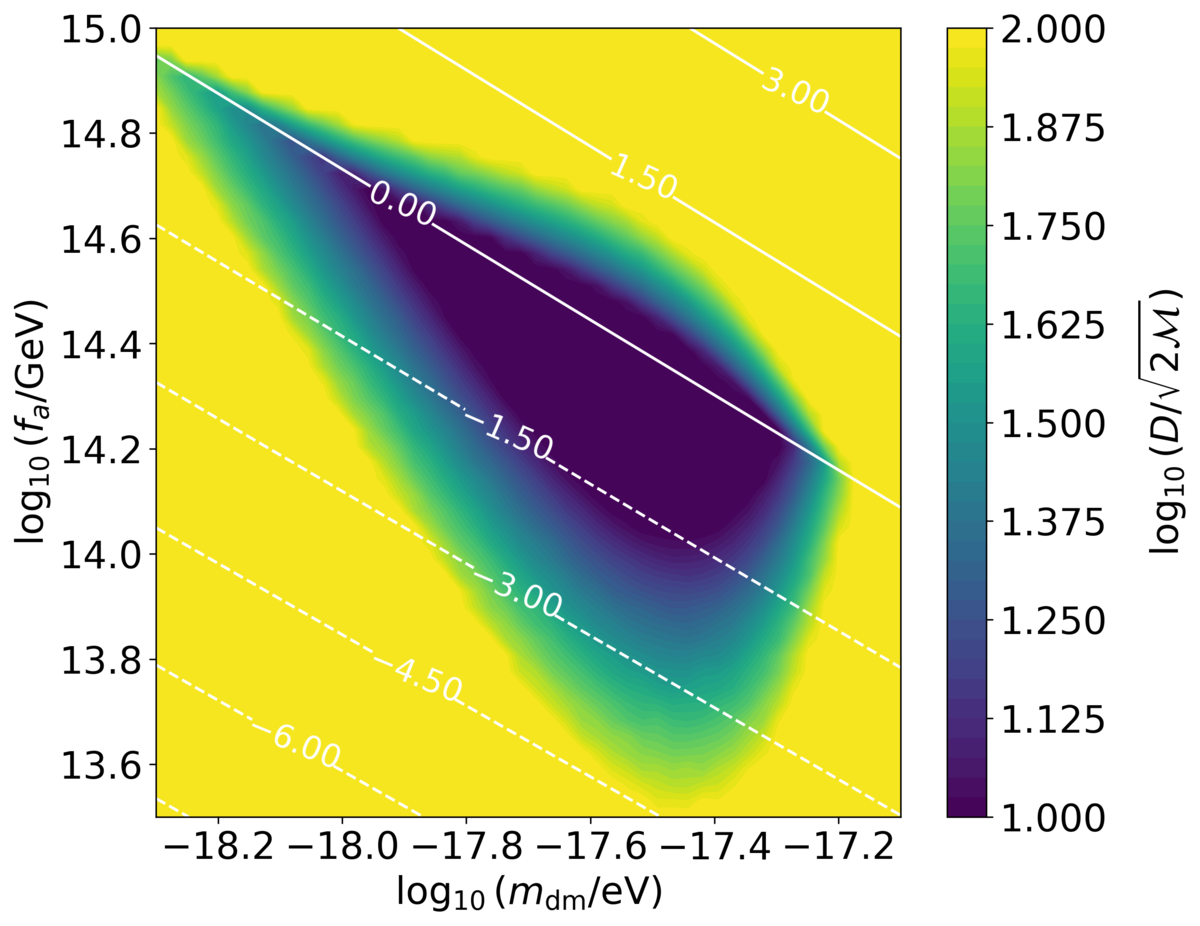}
    \end{subfigure}
    \caption{
    Contour maps in the $(m_{dm}, f_a)$ parameter space quantifying the distinguishability of GA-induced dephasing in the waveforms. Each panel corresponds to a different choice of the primary mass $M_{BH}$ and $\rho_{dm}$. We show:  $\{M_{BH} / M_\odot, \rho_{dm} / (\text{GeV/cm}^3)\}$: (upper left) $\{10^4, 10^4\}$, (upper right) $\{10^5, 10^5\}$,  (lower left) $\{10^6, 10^6\}$, and (lower right) $\{10^7, 10^7\}$. The SNR is shown in the range $\mathrm{SNR} \sim 20$--$100$, consistent with the expected values for EMRIs and IMRIs observable by LISA. Values below $10$ are disregarded (no detection), while values above $100$ are capped. For higher central BH mass, higher $\rho_{dm}$ are needed in order to find sufficient dephasing. High values indicate that the DM effect is weak and a large SNR would be required to distinguish the waveform from the background DM case, while low values indicate strong dephasing detectable at modest SNR. White lines denote the quantity $\log(\tau_{crit}/t_0)$ The white line at $\log(\tau_{crit}/t_0) = 0$ separates the regions between evolving and saturated GAs. The region forms a triangular shape. These are explained by the dependence on $m_{dm}$ of the density profile from Eq.~\ref{eq:density_profile} and the formation timescales involved for the whole combination of parameters. 
    }
    \label{fig:dm_contours}
\end{figure*}

\begin{figure}
    \includegraphics[width=1\linewidth]{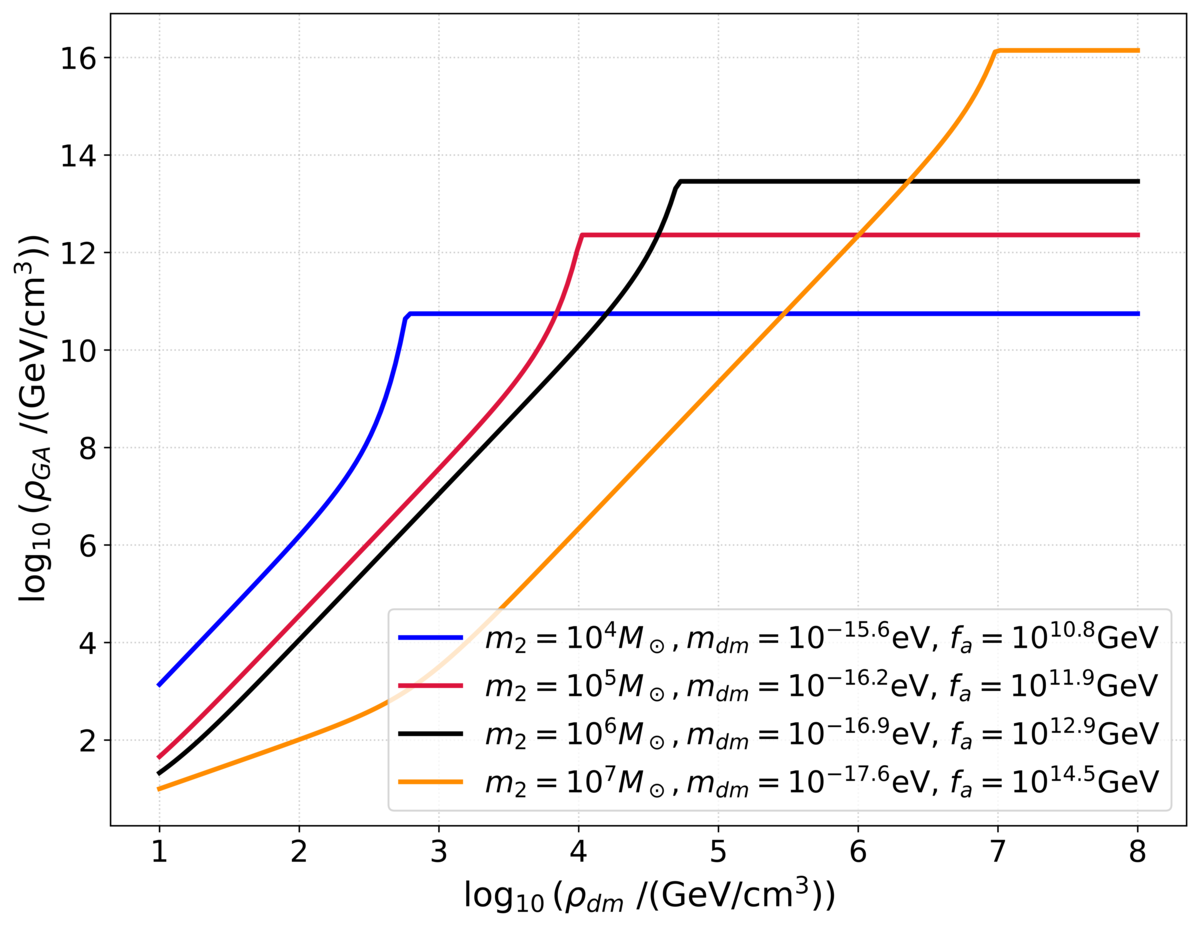}
   \caption{Relative change in the GA density at the companion’s initial position $r(10^{-4}\,\mathrm{Hz})$ with respect to $\rho_{\rm dm}$, for different central BH masses $M_{BH}$. The parameters $f_a, m_{dm}$ for the different BH masses are chosen such that they yield the maximum dephasing. The curves reach a plateau at $\rho_{\rm GA} = \rho_{\rm crit}$}    \label{fig:densities_densities}
\end{figure}

\begin{figure}
    \centering
    \includegraphics[width=1\linewidth]{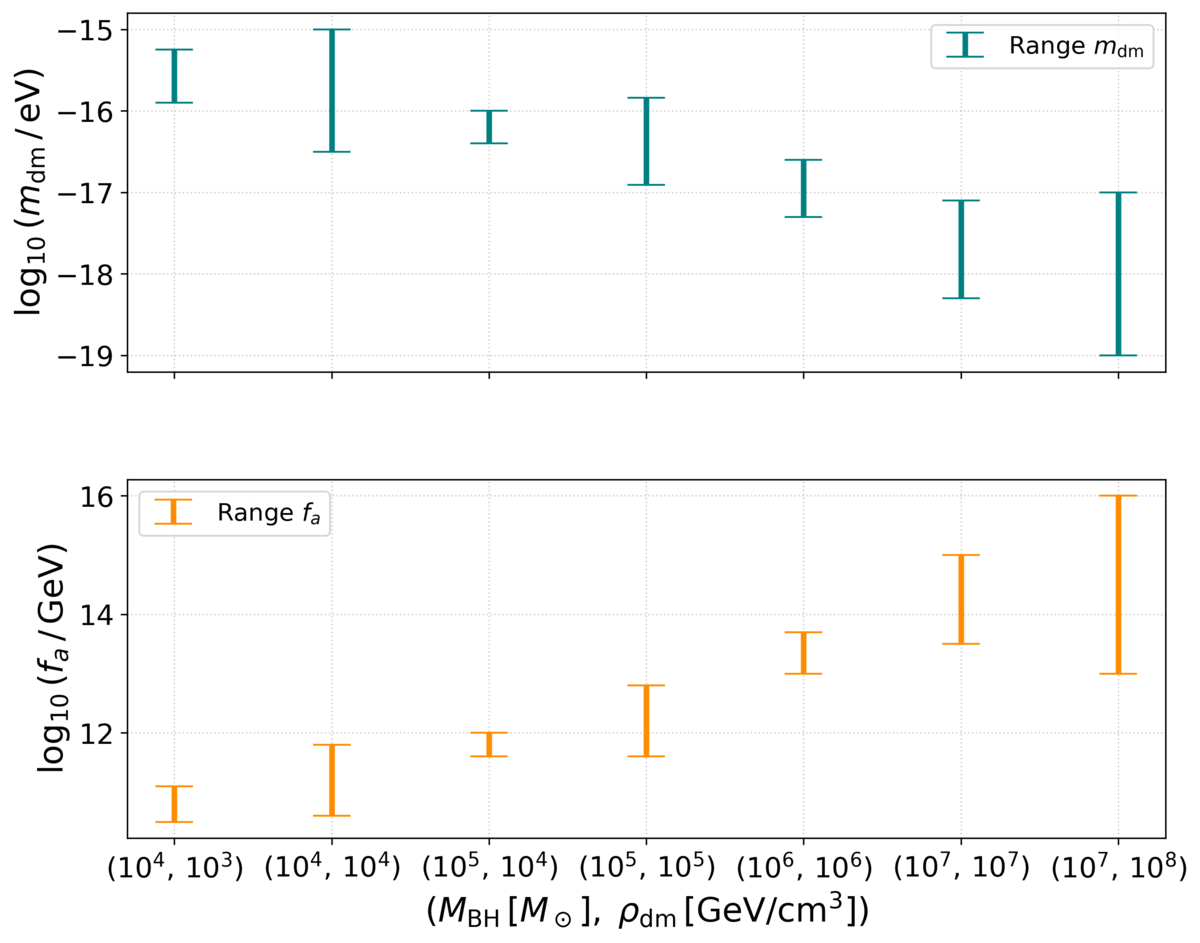}
    \caption{Ranges for $m_{dm}$ (upper plot) and $f_a$ (lower plot) that which can be constrained for different BH masses $M_{BH}$ and background DM densities $\rho_{dm}$. These include three conservative values for ${\rho_{dm} \sim (10^3-10^5) \,\rm{GeV}/cm^3}$ for the smaller pair of masses we explore $M_{BH} = 10^4M_{\odot}$ and $M_{BH} = 10^5M_{\odot}$, an more speculative values of $\rho_{dm} \sim (10^6-10^8) GeV/cm^3$ to the pair of more massive BHs we explore $M_{BH} = 10^6M_{\odot}$ and $M_{BH} = 10^7M_{\odot}$. We take the minimum and maximum values for each parameter $f_a$ and $m_{dm}$ as the minimum and maximum values which produce a dephasing detectable with $\mathrm{SNR} \leq 100$.}
    \label{fig:ranges}
\end{figure}

Fig.~\ref{fig:dm_contours} shows the minimum SNR required to distinguish a signal embedded in an ALP cloud from its vacuum counterpart across the ALP parameter space $(f_a, m_{dm})$ for BH with masses $M_{BH}=10^4M_\odot$ (first row), $M_{BH}=10^5M_\odot$ (second row) and $M_{BH}=(10^6 -10^7) M_\odot$ (third row left and right). The SNR is shown in the range $\mathrm{SNR} \sim 10$--$100$, consistent with expectations for EMRIs and IMRIs observable by LISA \cite{LISA_waveforms_2023}. We set a detectability threshold $\text{SNR}_{th}=10$, such that lower values are truncated. Thus, we have found that the environmental effects generated by the cloud are distinguishable from vacuum evolution across all binary configurations considered in this work for signal-to-noise ratios $\text{SNR} \geq 10$, corresponding to DM densities of $\rho_{\rm dm}=10^3 \,\rm GeV/cm^3$ (first row), $\rho_{\rm dm}=10^4 \,\rm GeV/cm^3$ (second row), and $\rho_{\rm dm}=(10^5 -10^7)\rm GeV/cm^3$ (third row, left and right panels). We observe a minimum SNR contour region with a triangular shape consistent with the maximum phase deviation shown in Fig.~\ref{fig:dm_contours_phi}. This is the region where the forming or saturated GAs have achieved the maximum densities permitted by the formation mechanism and its range in ALP masses is consistent with the halo density profile given in Eq.~\ref{eq:density_profile}, which exhibits two distinct regimes: $a_0 \ll r(f_{\rm GW})$ and $a_0 \gg r(f_{\rm GW})$, depending on the value of $m_{\rm dm}$ and the position of the orbiting BH, as illustrated in Fig.~\ref{fig:rho_mdm}. The white contour lines mark regions of different values for, $\tau_{\rm crit}/t_{\rm 0}$. As $m_{\rm dm}$ and $f_a$ increase, $\tau_{\rm crit}$ generally becomes larger, consistent with Eq. \eqref{eq:tcrit}, shifting the contours toward longer saturation times.
Moreover, higher values of $\rho_{dm}$ are required as increases. In Fig.~\ref{fig:densities_densities} we explain this behaviour by plotting the relative change in the GA density at the companion’s initial position $r(10^{-4}\,\mathrm{Hz})$ with respect to $\rho_{\rm dm}$, for different central BH masses $M_{BH}$. The overdensities increase with the background local density and reach a plateau at $\rho_{\rm GA} = \rho_{\rm crit}$. This is because orbiting companions in EMRIs with more massive central BHs have orbital radii larger than the radii at which we should find the peak of the GA density, which should be located at the center $r = 0$ (in practice truncated at the horizon). On the other hand, as $M_{BH}$ increases, $m_{dm}$ must decrease in order to satisfy the requirement $a_0 \gg 2m$. To achieve sufficiently high densities, the critical density Eq. \eqref{eq:critrho} requires $f_a$ to increase accordingly, and since $\Gamma$ (Eq.\eqref{eq:gamma}) is more sensitive to changes in $f_a$ than in $m_{dm}$, the background density $\rho_{dm}$ must compensate by increasing as well. The plots reveal a degeneracy between the model parameters, consistent with the analytical structure of Eqs.\eqref{eq:gamma} and \eqref{eq:critrho}: different combinations of $m_{dm}$ and $f_a$ can yield identical values of both the formation timescale and the maximum achievable GA density. 
In Fig.~\ref{fig:ranges}, we show the DM particle parameter range which can be potentially constrained by this model. We take the minimum and maximum values for each parameter $f_a$ and $m_{dm}$ as the minimum and maximum values which produce a dephasing detectable with $\text{SNR} \leq 100$. GAs generated during cosmological times around different EMRI configurations can probe $m_{dm} \sim  10^{-15}$--$10^{-19}$\,eV and $f_a \sim 10^{11}$--$ 10^{17}$ GeV. This calculation includes both conservative background DM densities and higher density configurations. Cases for $M_{BH} > 10^7M_{\odot}$ are not shown but can be computed analogously. 
\subsection{PARAMETER ESTIMATION}

To quantify the precision with which the ALP parameters $(f_a, m_{\mathrm{dm}})$ can be constrained, we performed a Bayesian parameter estimation analysis by sampling the posterior distribution from the Fisher-Matrix inspired likelihood~\cite{JimenezForteza:2018rwr,Chatziioannou:2017tdw}
\begin{equation}
    \log\mathcal{L} = -\text{SNR}^2\,\mathcal{M}\left[h_{GA}(f_a^{\mathrm{fid}}, 
    m_{\mathrm{dm}}^{\mathrm{fid}})\,\big|\,h_{GA}(f_a, m_{\mathrm{dm}})\right].
    \label{eq:logL_Fisher}
\end{equation}
where the mismatch $\mathcal{M}$ is computed between a waveform generated at the fiducial injection values and a template evaluated at $(f_a, m_\mathrm{dm})$. We adopt flat priors over the logarithmic parameter space consistent with the ranges in Fig.~\ref{fig:dm_contours} and inject signals at SNR $= 20$, corresponding to the minimum detection threshold for EMRIs with LISA~\cite{LISA_waveforms_2023}.

\begin{figure*}[ht]
    \centering
    \begin{subfigure}{0.45\textwidth}
        \centering
        \includegraphics[width=\linewidth]{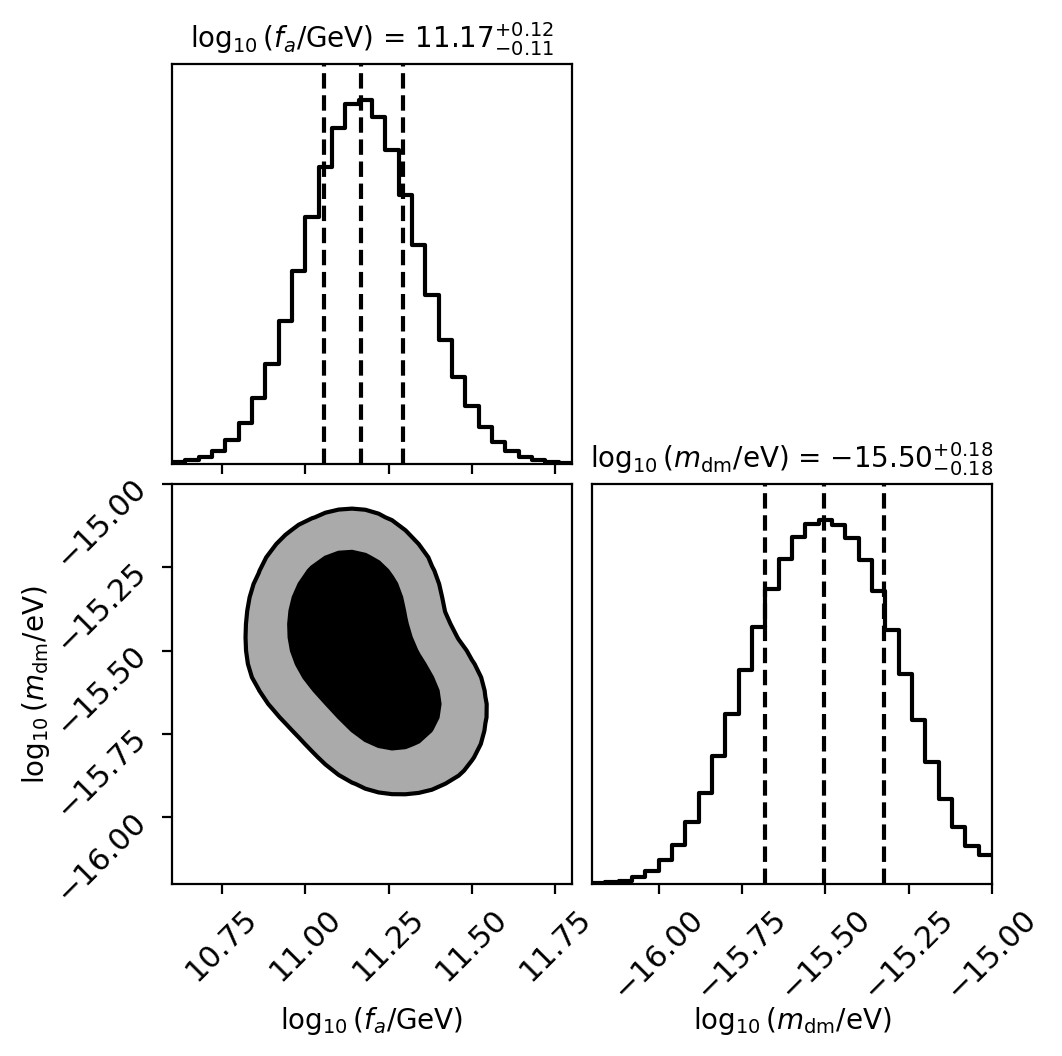}
    \end{subfigure}
    \hfill
    \begin{subfigure}{0.45\textwidth}
        \centering
        \includegraphics[width=\linewidth]{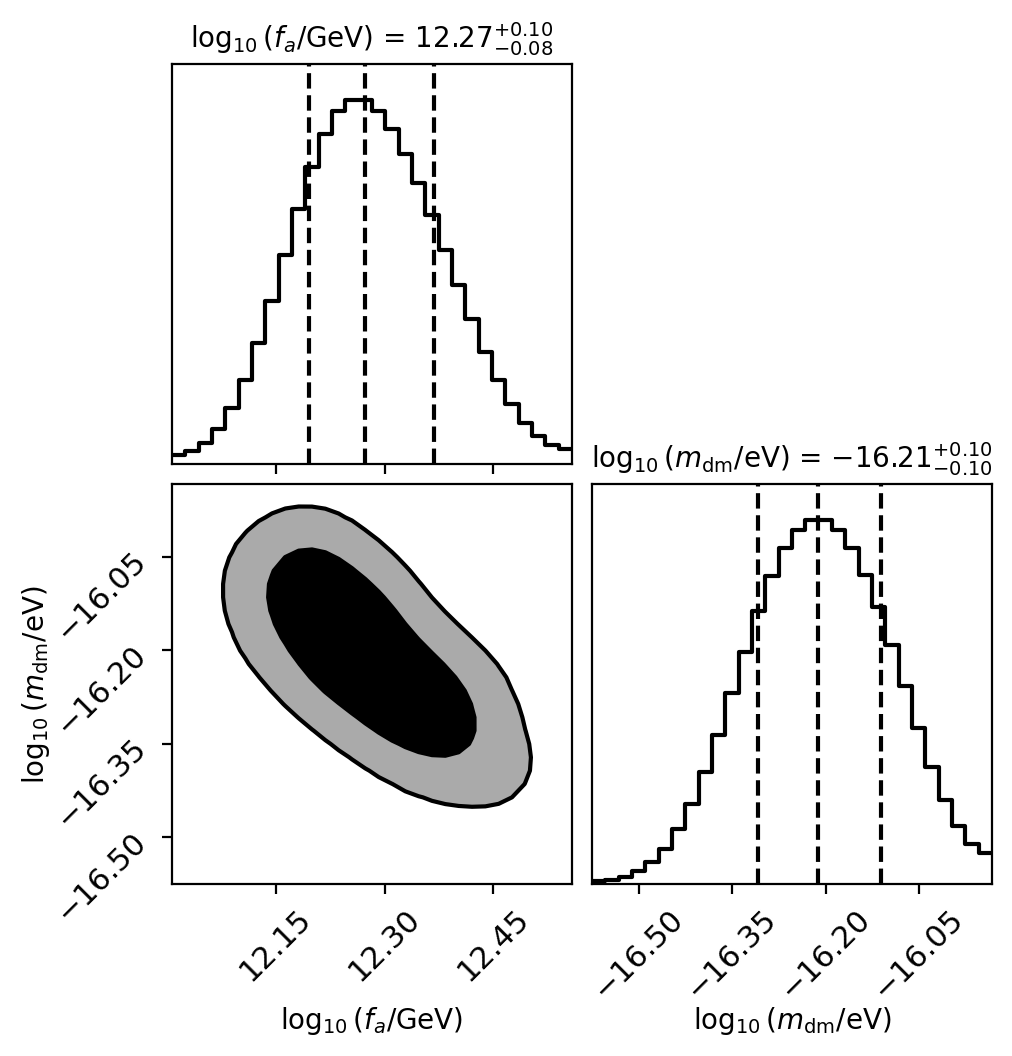}
    \end{subfigure}

    \caption{Posterior distributions obtained from the likelihood
    $\mathcal{L} = -\mathrm{SNR}^2 \,\mathcal{M}$ comparing dynamical-DM
    waveforms with injected signals with a conservative $\text{SNR} = 20$. Each panel corresponds to a different binary configuration and DM density, $M_{BH} = 10^4 M_{\odot}$, $\rho_{dm} = 10^4 GeV/cm^3$ (left) and $M_{BH} = 10^5 M_{\odot}$, $\rho_{dm} = 10^5 GeV/cm^3$ (right). The off-diagonal panel shows the joint posterior, with filled contours
enclosing the 68\% and 95\% credible regions. The diagonal panels show the marginalized one-dimensional posteriors, with dashed vertical lines indicating the 16th, 50th, and 84th percentiles. Titles report the median and the $1\sigma$ credible interval for each parameter. The joint recovery with four parameters of both the particle and the binary masses does not alter these results due to the PN order at which the DF effect dominates.}
    \label{fig:corner_plot}
\end{figure*}

In Fig.~\ref{fig:corner_plot} we show the resulting posterior distributions for two representative injection points chosen to lie near the region of maximum mismatch in each configuration, as it can be inferred from Fig.~\ref{fig:dm_contours}. Both the injected and recovered parameters are organized in Table \ref{tab:parameters}. In both cases the posterior peaks at the injected values, confirming accurate parameter recovery. We found a degeneracy between the parameters, mainly arising from the structure of Eqs.~\ref{eq:gamma} and \ref{eq:critrho}, which is also reflected in the distinguishability contours shown in Fig.~\ref{fig:dm_contours}.
\begin{table*}[t]
\centering
\renewcommand{\arraystretch}{1.3}
\setlength{\tabcolsep}{10pt}
\begin{tabular}{|l|c|c|}
\hline
$(M_{BH}, \rho_{dm})$
 & ($10^4M_{\odot}$,$10^3\,\mathrm{GeV/cm}^3$)  
 & ($10^5M_{\odot}$,$10^4\,\mathrm{GeV/cm}^3$)  \\
\hline
$f_a^{\mathrm{fid}}$ [GeV] 
& $1.585 \times 10^{11}$ 
& $1.995 \times 10^{12}$ \\[4pt]
$m_{\mathrm{dm}}^{\mathrm{fid}}$ [eV] 
& $3.162 \times 10^{-16}$ 
& $5.888 \times 10^{-17}$ \\
\hline
$f_a$ [GeV] 
& $\left(1.479^{+0.471}_{-0.331}\right) \times 10^{11}$ 
& $\left(1.862^{+0.486}_{-0.313}\right) \times 10^{12}$ \\[6pt]
$m_{\mathrm{dm}}$ [eV] 
& $\left(3.162^{+1.624}_{-1.073}\right) \times 10^{-16}$ 
& $\left(6.166^{+1.597}_{-1.268}\right) \times 10^{-17}$ \\
\hline
\end{tabular}
\caption{Fiducial and recovered parameter values for the two benchmark configurations.}
\label{tab:parameters}
\end{table*}
The uncertainty on $f_{\mathrm{a}}$ ranges from $-22.0\%$ to 
$+24.0\%$, while $m_{\mathrm{dm}}$ is constrained within $-34.6\%$ / $+42.7\%$. 
For the second configuration the $1\sigma$ credible intervals reach $-17.4\%$ / 
$+24.5\%$ for $f_{\mathrm{a}}$, and $-20.9\%$ / $+26.3\%$ for $m_{\mathrm{dm}}$.
We have focused only on performing parameter estimation in the ALP parameters because they are completely uncorrelated with $\mathcal{M}_c$, which can be recovered from the quadrupolar vacuum evolution term in Eq.~\ref{eq:ODE}. This same argument applies to any standard effect with a non-negative PN order, such as spin-orbit coupling (1.5\,PN), spin-spin interactions (2\,PN), gravitational self-force corrections at 1\,PN and higher, or tidal deformability at 5\,PN \cite{blanchet2024}. Introducing eccentricity would add higher harmonics at $f_n = n f_{\rm orb}$. At leading adiabatic order and for quasi-circular inspirals, the cloud-induced
dephasing of the $(\ell,m)$ harmonic scales linearly with the azimuthal number, $\delta \Phi_{\ell m} \simeq \frac{m}{2}\,\delta \Phi_{22}$ since all harmonics are sourced by the same underlying correction to the orbital phase. Therefore higher modes beyond $m=2$ would add phase information but is not expected to qualitatively alter the environmental effects signature. 

\begin{figure*}[ht]
    \centering
    \begin{subfigure}{0.47\textwidth}
        \centering
        \includegraphics[width=\linewidth]{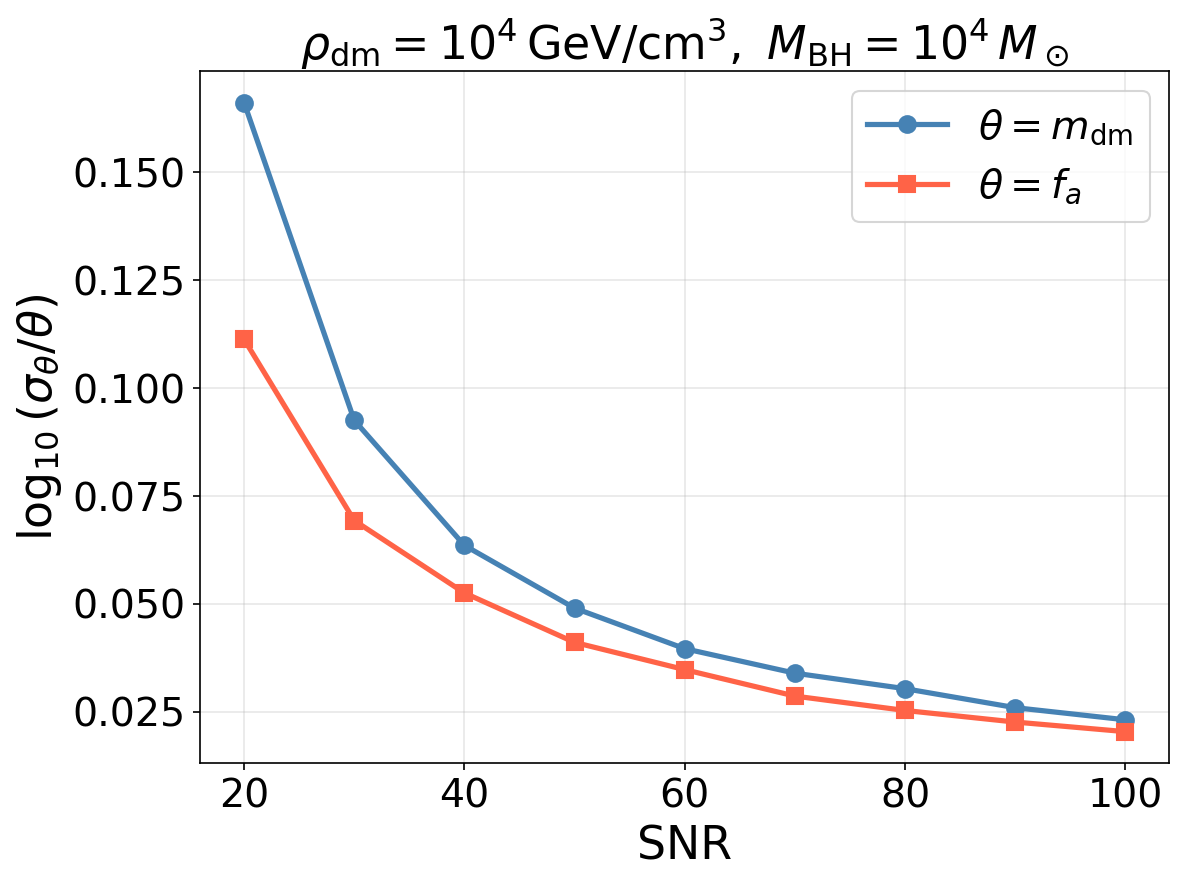}
    \end{subfigure}
    \hfill
    \begin{subfigure}{0.47\textwidth}
        \centering
        \includegraphics[width=\linewidth]{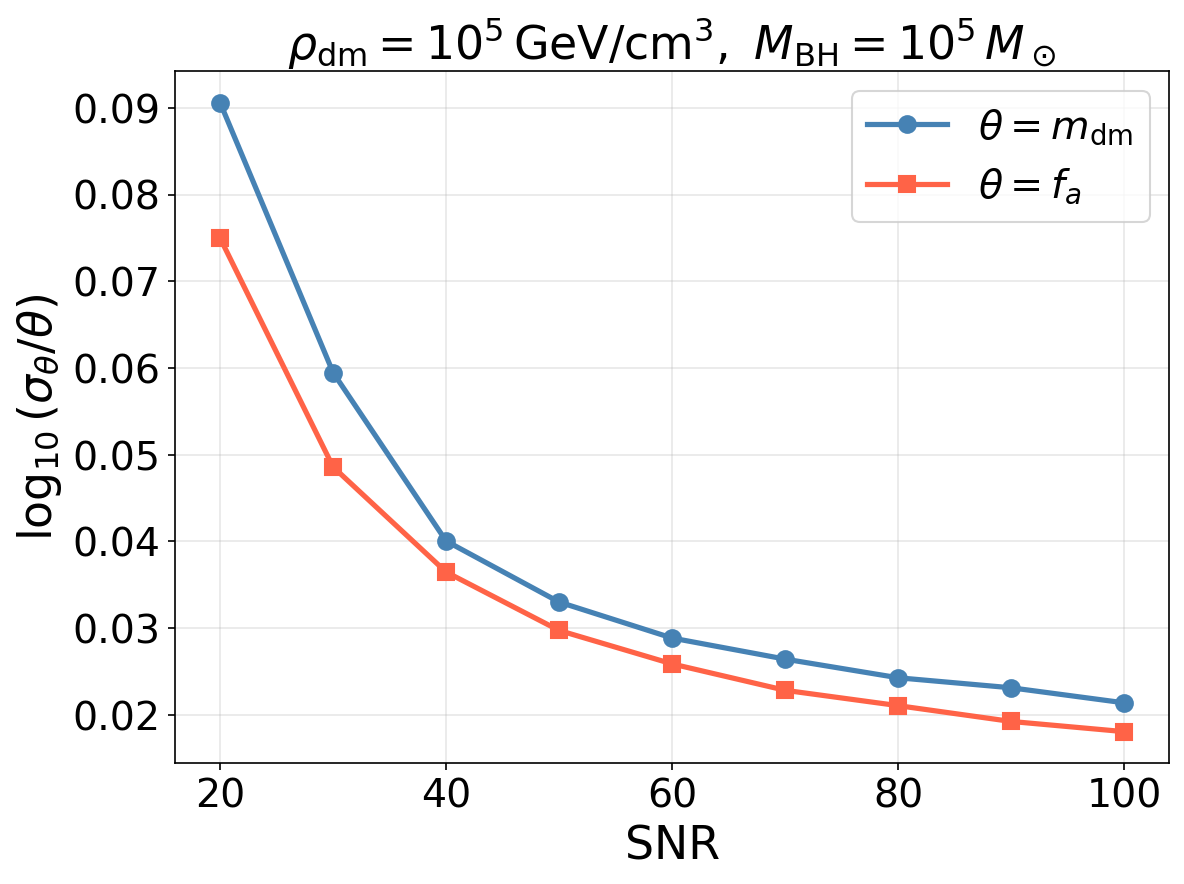}
    \end{subfigure}

    \caption{Fractional parameter uncertainties $\sigma_{f_a}/f_a$ and $\sigma_{m_{dm}}/m_{dm}$ as a function of the SNR for different system configurations from one-dimensional posterior distributions for the injected values $\log_{10}(f_a) = 11.2$ $\log_{10}(m_{dm}) = -15.5$ for $M_{BH} = 10^4M_{\odot}$, $\rho_{dm} = 10^4 GeV/cm^3$ (left) and $\log_{10}(f_a) = 12.3$ $\log_{10}(m_{dm}) = -16.2$ for $M_{BH} = 10^5M_{\odot}$, $\rho_{dm} = 10^5 GeV/cm^3$ (right). As expected from Fisher-matrix theory, the fractional uncertainties decrease monotonically with increasing SNR, approximately following a $\mathrm{SNR}^{-1}$ scaling.}
    \label{fig:sigma_SNR}
\end{figure*}

In Fig.~\ref{fig:sigma_SNR} we have calculated the fractional parameter uncertainties $\sigma_{\theta}/\theta$, for each parameter $\theta = f_a, m_{\rm dm}$ and two representative injection points chosen to lie near the region of maximum mismatch, as a function of SNR $\rho$. They decrease monotonically with increasing SNR for both parameters, approximately following a $\mathrm{SNR}^{-1}$ scaling consistent with Fisher-matrix expectations. For $\mathrm{SNR} = 20$, we find fractional uncertainties at the level of $\sigma_{m_\mathrm{dm}}/m_\mathrm{dm}, \sigma_{f_a}/f_a \gtrsim 10\%$ for the first binary configuration and $\sigma_{m_\mathrm{dm}}/m_\mathrm{dm}, \sigma_{f_a}/f_a \lesssim 10\%$ for the second one. At $\mathrm{SNR} \sim 80$, these improve to $\sim 2\% -3\%$ for both configurations. These results show that both parameters can be recovered with small uncertainties for the expected LISA SNR range between $20 - 100$.

\begin{figure*}[h!]
    \centering
    \begin{subfigure}{0.45\textwidth}
        \centering
        \includegraphics[width=\linewidth]{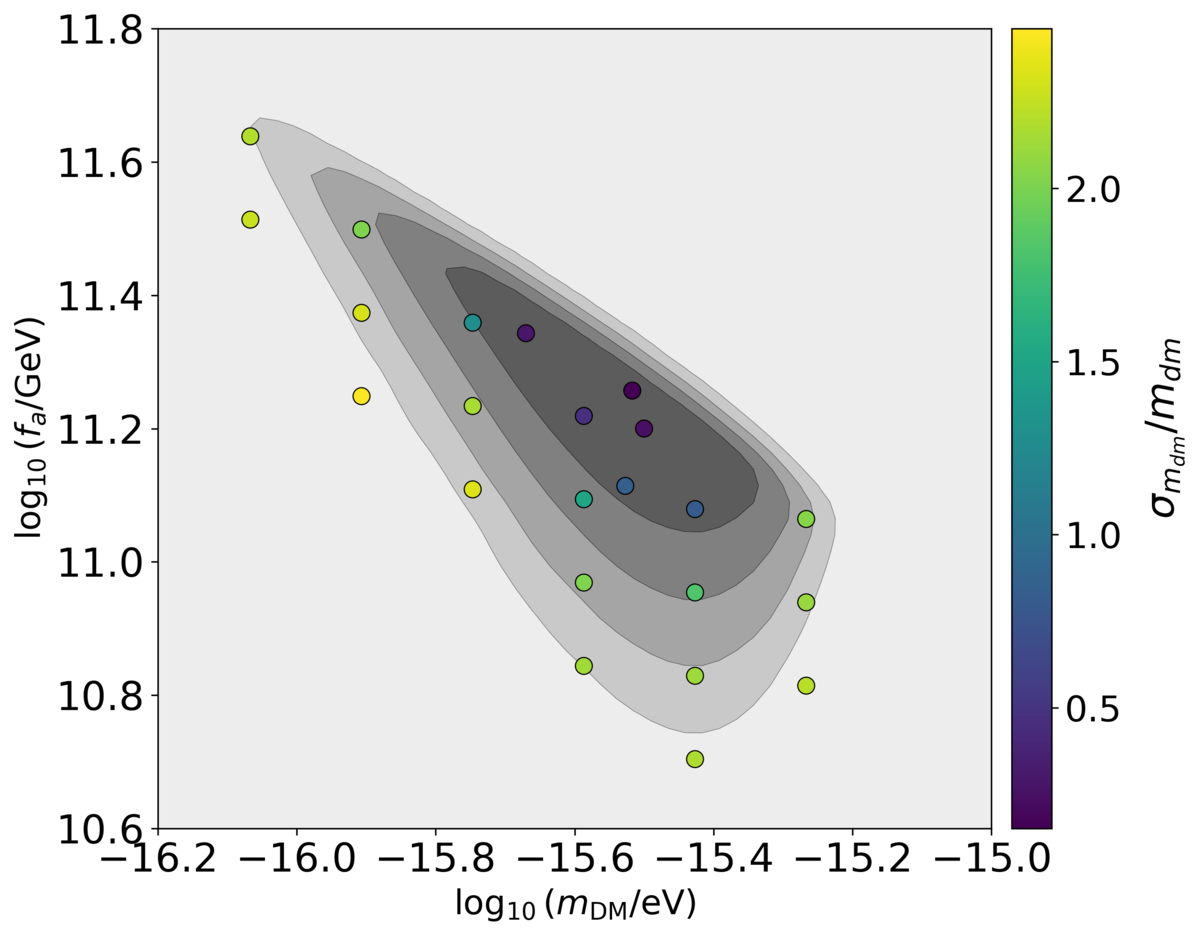}
    \end{subfigure}
    \hfill
    \begin{subfigure}{0.45\textwidth}
        \centering
        \includegraphics[width=\linewidth]{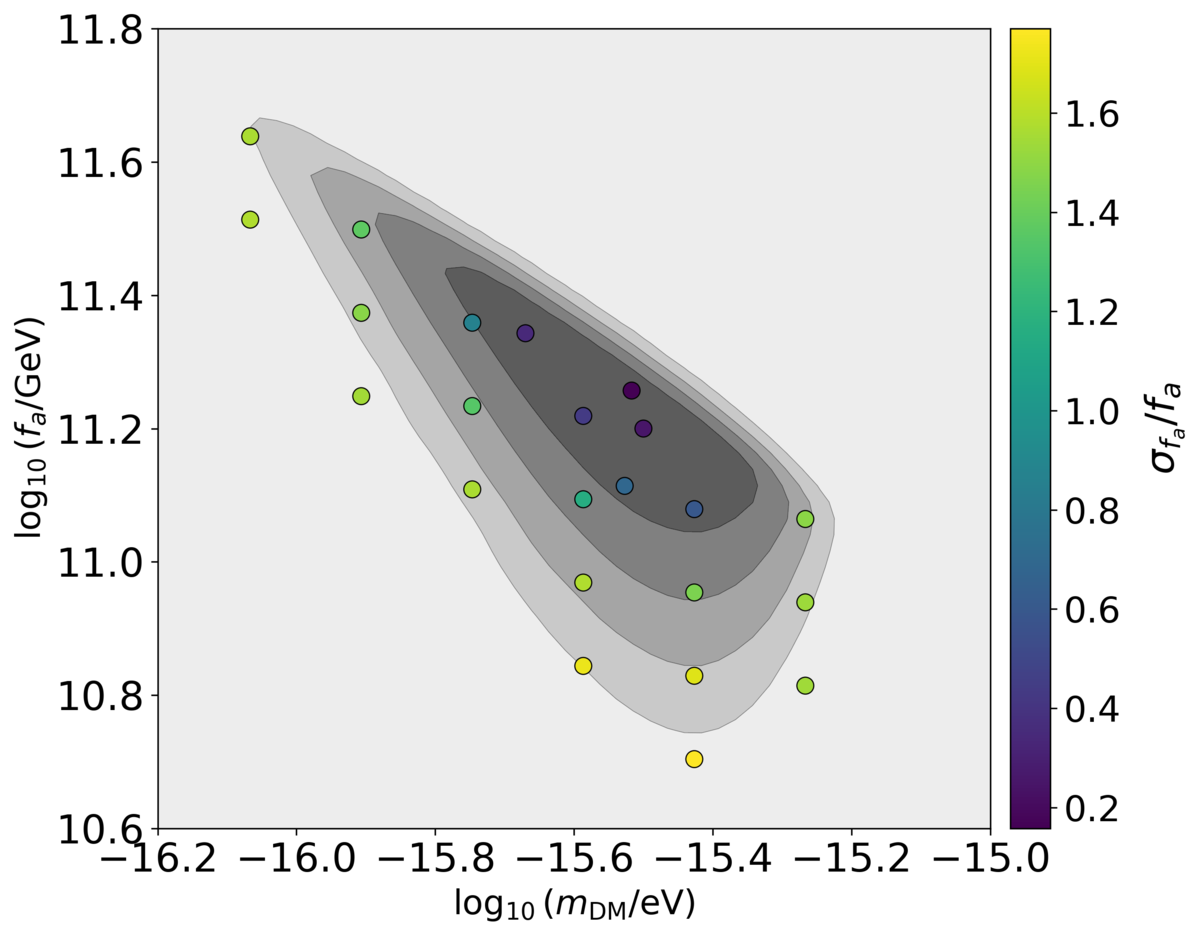}
    \end{subfigure}

    \vspace{0.5cm} 

    \begin{subfigure}{0.45\textwidth}
        \centering
        \includegraphics[width=\linewidth]{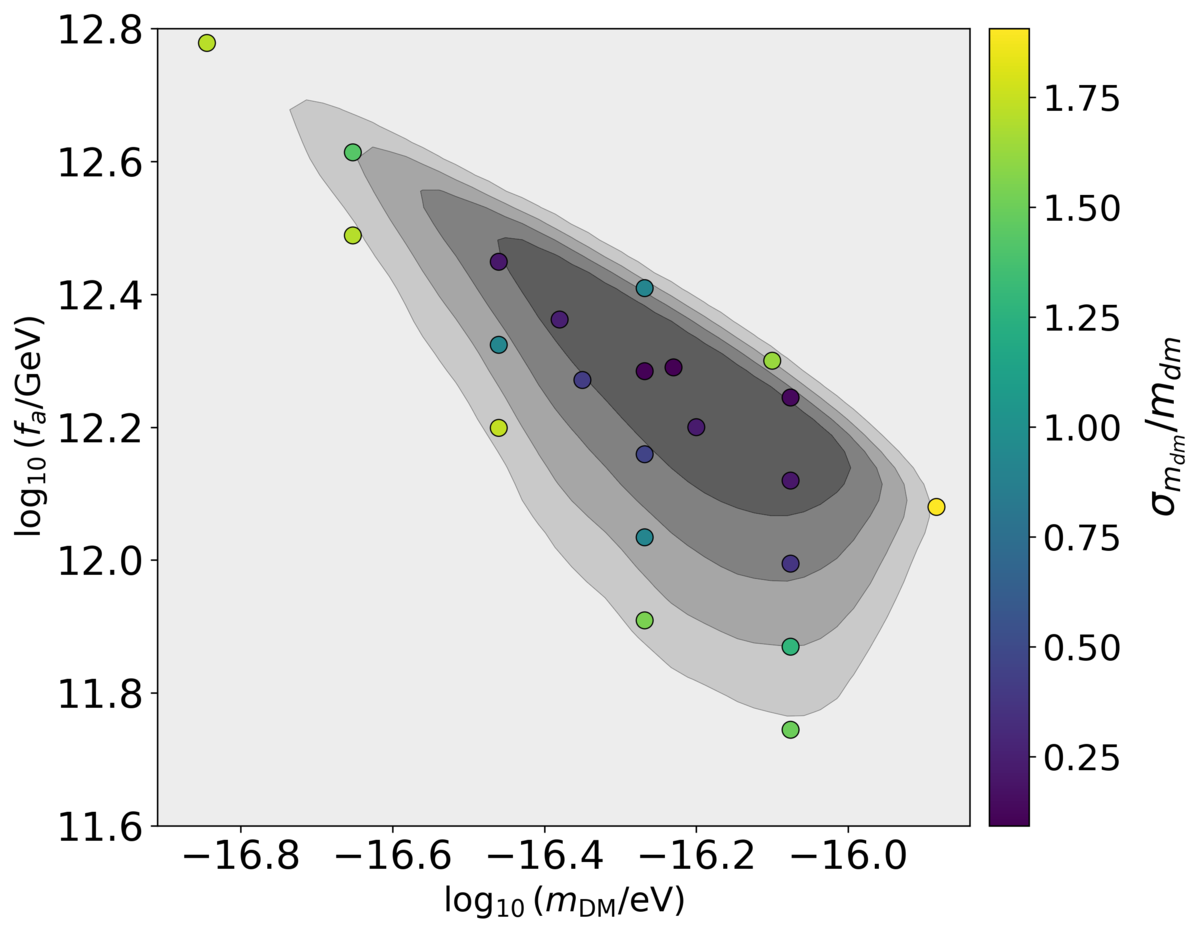}
    \end{subfigure}
    \hfill
    \begin{subfigure}{0.45\textwidth}
        \centering
        \includegraphics[width=\linewidth]{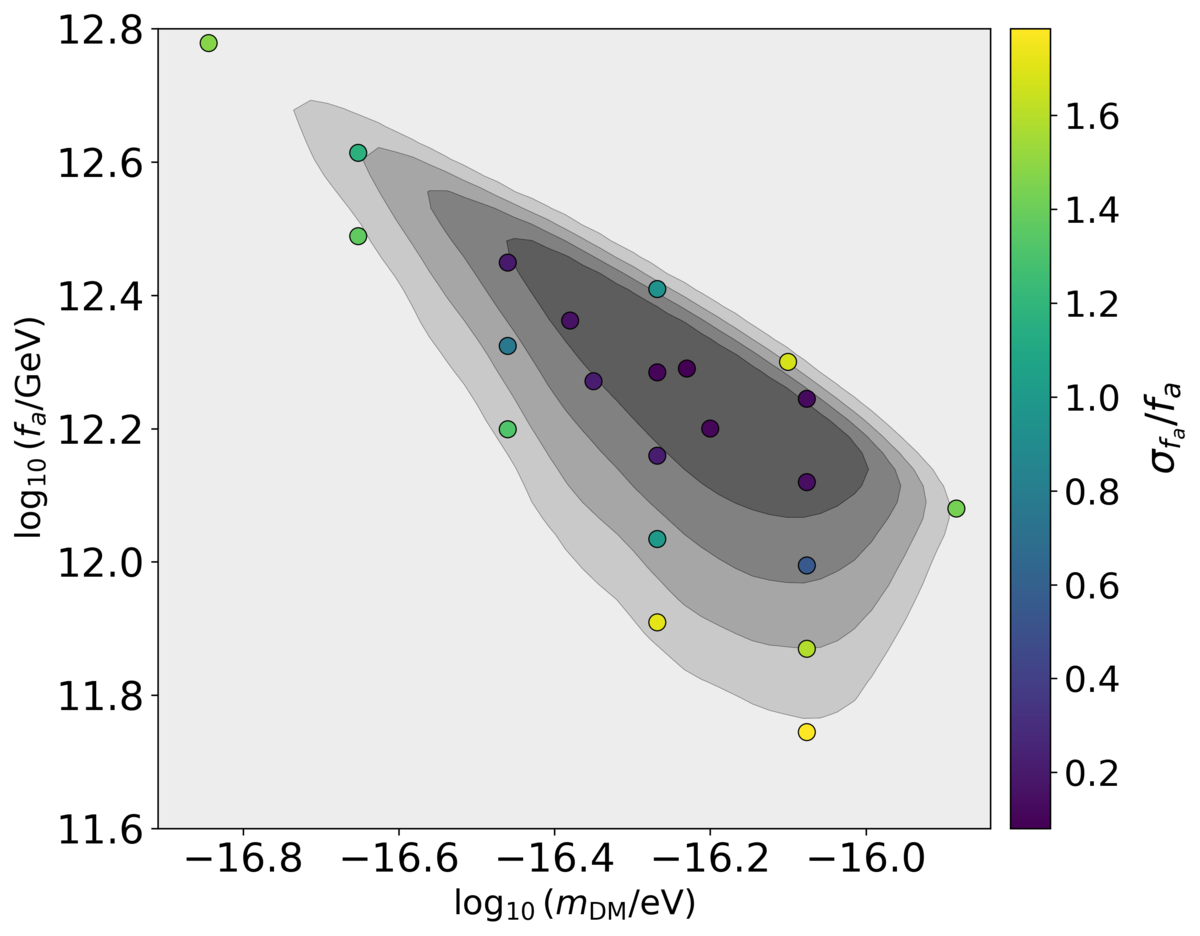}
    \end{subfigure}

    \caption{Fractional parameter uncertainties  for the DM halo 
parameters $m_\mathrm{dm}$ (left panels) and $f_a$ (right panels), obtained from one-dimensional posterior distributions at each injection point in the $(m_\mathrm{dm}, f_a)$ parameter space with $SNR = 20$. These are for $M_{BH} = 10^4 M_{\odot}$ with $\rho_{dm} = 10^4 GeV/cm^3$ for the upper plots and $M_{BH} = 10^5 M_{\odot}$ with $\rho_{dm} =  10^5 GeV/cm^3$ for the lower plots. The parameter ranges of $f_a$ and $m_{dm}$ are chosen to coincide with the regions of detectable mismatch between waveforms produced by binaries embedded in the background DM density and in a GA, which are marked in gray. Lighter (yellow) colors indicating better parameter recovery. Points closer to the region of high mismatch (or less needed SNR to be detectable as in Fig.~\ref{fig:dm_contours}) show smaller fractional uncertainties, consistent with the expectation that better-constrained parameters correspond to stronger DM induced dephasing. For both configurations, fractional uncertainties range from $\sim 20\%$--$80\%$ in the high mismatch region up to $\sim 150\%$--$200\%$ near the boundaries, for both $m_{\mathrm{dm}}$ and $f_a$. Towards these boundaries, the parameters become increasingly degenerate, leading to reduced robustness in 
the inferred uncertainties.}
\label{fig:sigma_map}
\end{figure*}

In Fig.~\ref{fig:sigma_map} we show the spatial distribution of $\sigma_{\theta_i}/\theta_i$ across the $(m_\mathrm{dm}, f_a)$ parameter space for a fixed $\mathrm{SNR} = 20$. These were calculated by recovering different parameter injections in $\log \mathcal{L}$. By marking in gray the region where we have computed the mismatch $\mathcal{M}(h_{BG}|h_{GA})$, where $h_v$ is the strain in vacuum and $h_{GA}$ is the strain for EMRIs embedded in a GA, the fractional uncertainties are smallest in the region of highest mismatch, consistent with the fact that the regions where the ALP parameters can be most precisely constrained correspond to those producing the strongest environmental effect, and thus the likelihood from Eq.~\ref{eq:logL} is most sharply peaked. 

Specifically, within the most sensitive regions of the parameter space (yellow points), the fractional uncertainties reach values as low as $\sigma_i/\theta_i \approx 0.05\%-0.3\%$, corresponding to $\sim 5\%$--$30\%$. Toward the periphery of the dephasing region (dark purple points), the fractional uncertainties increase to $\sigma_X/X \sim 1.5$--$3.0$, indicating that parameter recovery becomes significantly degraded. We note that near the boundary of the detectable region, where the mismatch approaches zero, the posteriors broaden and the parameters become strongly correlated. Consequently, parameter estimation should be treated with caution outside the gray contour region, as the inherent degeneracies in those regimes prevent a robust recovery.
\section{Discussion}
In this work, we have investigated the impact of ALP self-interactions on the formation and evolution of GA surrounding massive BHs in EMRIs and IMRIs, and assessed their potential imprint on GW observations in the LISA band. By incorporating a realistic dynamical formation mechanism for bosonic halos~\cite{budker2023genericformationmechanismultralight}  and modeling their effect on binary evolution, we have shown that these environments can induce observable dephasing in the GW signal when the cloud becomes sufficiently dense and long-lived.  

We explored ultralight boson masses $m_{\rm dm} \lesssim 10^{-14}\,\mathrm{eV}$, with self-interactions parametrized by the ALP decay constant $f_a$, sourced by EMRIs and IMRIs with a central BH with mass in the range $10^4$--$10^7\,M_\odot$. Therefore, we have studied the effects of the cloud on the phase evolution of the small object and compare them to the EMRI/IMRI scenario in the DM background. 

We have found that large values of $m_{\rm dm}$ lead to the formation of clouds with small Bohr radii, for which DF becomes inefficient and the orbit of the secondary remains largely unaffected. Conversely, smaller values of $m_{\rm dm}$ produce more extended clouds, but with lower densities ($\rho \sim a_0^{-3}$), resulting again in a weak impact on the orbital evolution. As a consequence, only an intermediate range of $m_{\rm dm}$ and $f_a$ leads to significant dephasing in the GW signal. 

On the other hand, the set of values of $f_a$, given a particular range of boson masses, sets the timescale and critical density at which the GAs form, such that the values of the decay constants explored here are such so the timescales involved in order to achieve sufficiently high densities are of the order of the age of the universe. 

Our results show that LISA could probe boson masses in the approximate range $m_{\rm dm} \sim (10^{-17}$--$ 10^{-15})\,\mathrm{eV}$ and decay constants $f_a \sim (3 \times 10^{10}$--$6 \times 10^{12})\,\mathrm{GeV}$ under conservative assumptions for the ambient DM density in the range $\rho_{dm} \sim (10^3-10^4)GeV/cm^3$ and central BH masses of $M_{BH}\sim (10^4$--$10^5\,M_\odot)$. Using a Fisher Matrix framework, we have studied the fractional uncertainties $\sigma_{\theta_i}/\theta_i$ change for both $m_{\mathrm{dm}}$. For $\mathrm{SNR} = 20$ these range from less than $\sim 10\%$ in the high-mismatch region to $\sim 50\%$--$80\%$. In both configurations, points lying closer to the region of high mismatch exhibit systematically smaller fractional uncertainties, consistent with the expectation that stronger DM induced dephasing leads to tighter posterior distributions, and become increasingly degenerate in the boundaries of this region. For increasing $\mathrm{SNR}$ these uncertainties decrease monotonically, approximately following a $\mathrm{SNR}^{-1}$ scaling. For $\mathrm{SNR} \gtrsim 80$ and masses $M_{BH} = 10^4 M_\odot$ and $M_{BH} = 10^5 M_\odot$, the fractional uncertainties on the halo parameters improve to near the percent.

Allowing for background density orders of magnitude higher, $\rho_{dm} \sim (10^6-10^7)GeV/cm^3$, we show that the range of parameters over which we can impose constraints gets broader, up to $f_a \sim 10^{16}$ GeV and down to $m_{dm} \sim \times 10^{-19}$ eV. This combination of tiny masses and highly suppressed couplings situates the field well beyond the reach of laboratory-based searches, instead pointing to ALPs arising from high-energy symmetry breaking or compactification scenarios \cite{Arvanitaki_2010}, naturally interpreted as members of the broader "axiverse" predicted in many ultraviolet completions. Although the misalignment mechanism for a conventional axion potential underproduces the DM abundance for the plotted values of $m_{dm}$ and $f_a$, other scalar potentials or different production mechanisms could provide the correct abundance~\cite{budker2023genericformationmechanismultralight}. We note however that these constraints can be made model-independent by changing $f_a$ for $\lambda$, and that the particles are inferred solely through their gravitational influence, mirroring the only interaction for which DM currently has direct empirical support.  

Our results are broadly consistent with previous studies in the sense that self-interacting bosonic environments can produce measurable dephasing in the LISA band, but the physical regime considered here is different. Equilibrium and self-gravitating configurations, such as those studied in Refs.~\cite{Boudon_2024,banik2025bosonstarshostingblack,Chavanis:2019bnu}, can reach much larger central densities, and therefore often predict larger environmental effects. In our case, however, the cloud is not imposed as an equilibrium overdensity: it is generated dynamically from a realistic ambient halo density through the capture mechanism of Ref.~\cite{budker2023genericformationmechanismultralight}. As a result, the maximum cloud mass remains small, $\sim 10^{-6}M_{\rm BH}$, so neglecting the self-gravity of the cloud and treating the potential as dominated by the host BH is self-consistent. This also explains why the dephasings we obtain are smaller than in some equilibrium-cloud models, while remaining detectable by LISA in a well-defined region of parameter space.

An important extension of this work would be the inclusion of eccentric EMRI configurations. In this paper, we have restricted our analysis to quasi-circular inspirals, which provide a useful first approximation but do not capture the full range of astrophysically relevant scenarios. Eccentricity can significantly affect both the orbital evolution and the GW signal, potentially enhancing or modifying the impact of the boson cloud on the phase evolution. Another possible extension could be including the effects of DF on the higher harmonics of the GW signal, which enter the phase evolution in a manner analogous to the dominant $(2,2)$ mode. Specifically, for a given harmonic $(\ell,m)$, the phase contribution from DF can be expressed as a simple scaling of the dominant mode phase $\phi_{\ell m}^{\rm DF} = \frac{m}{2} \, \phi_{22}^{\rm DF}$, where $\phi_{22}^{\rm DF}$ denotes the DF-induced phase shift of the $(2,2)$ mode. This relation allows one to incorporate DF consistently across all relevant harmonics without recalculating the phase from scratch for each mode. 

\section{Acknowledgements}

We thank Joshua Eby, Laura Sberna and Gianfranco Bertone for insightful comments on this work.

This work was supported by the Universitat de les Illes Balears (UIB) with funds from the Programa de Foment de la Recerca i la Innovaci\'o de la UIB 2024-2026 (supported by the yearly plan of the Tourist Stay Tax ITS2023-086); Spanish Ministerio de Ciencia, Innovaci\'on y Universidades (Beatriz Galindo, BG22-00034),  the Spanish Agencia Estatal de Investigaci\'on grants PID2022-138626NB-I00, RED2024-153978-E, RED2024-153735-E, funded by MICIU/AEI/10.13039/501100011033 and the ERDF/EU; and the Comunitat Aut\`onoma de les Illes Balears through the Conselleria d'Educaci\'o i Universitats with funds from the ERDF (SINCO2022/18146 - Plataforma HiTech-IAC3-BIO). 

The authors thankfully acknowledge the computer resources at MareNostrum~5 and the technical support provided by the Barcelona Supercomputing Center (BSC) through grants No. RES-AECT-2025-3-0048, RES-AECT-2025-3-0011, RES-AECT-2025-3-0040 and from the Red Espa\~nola de Supercomputaci\'on (RES).

The authors gratefully acknowledge the computer resources at Artemisa, funded by the European Union and the Conselleria of Education, Research, Culture and Sports of the Generalitat Valenciana through the Project IDIFEDER/2018/048 of the Operative Programme FEDER 2014-2020 of the Comunitat Valenciana, with the financial support from the MCIU with funds from the European Union NextGenerationEU (PRTR-C17.I01) and Generalitat Valenciana (ASFAE/2022/024), as well as the technical support provided by the Instituto de F\'isica Corpuscular, IFIC (CSIC-UV). 

\section{Appendix}
\label{sec:appendix}
\subsection{Parameter estimation and assessment of detectability}
We compare two waveforms by computing their mismatch 
\begin{equation}
\label{eq:MM}
\mathcal{M}
\equiv 1 - 
\frac{\langle h_1 | h_2 \rangle}
{\sqrt{\langle h_1 | h_1 \rangle
       \langle h_2 | h_2 \rangle}}.
\end{equation}
The noise-weighted inner product is given by
\begin{equation}
\label{eq:dotp}
\langle h_1 | h_2 \rangle
= 4\,\mathrm{Re}\!\int_{f_{\rm min}}^{f_{\rm max}}
\frac{\tilde{h}_1(f)\,\tilde{h}_2^*(f)}{S_n(f)}\,\mathrm{d}f,
\end{equation}
with $\tilde{a}(f)$ denoting the Fourier transform of $a(t)$ and
$S_n(f)$ the one-sided LISA noise power spectral density \cite{Robson_2019}.
We define the SNR of a GW signal $h(t)$ as
\begin{equation}
\label{eq:SNR}
\mathrm{SNR}^2 \equiv \langle h \mid h \rangle.
\end{equation}
For a model described by parameters $\boldsymbol{\theta} = 
(\theta_1, \theta_2, \dots, \theta_n)$, with observed data $\mathbf{d}$,
and given the likelihood $\mathcal{L}(\boldsymbol{\theta}) = 
P(\mathbf{d} \,|\, \boldsymbol{\theta})$, we sample the posterior 
distribution defined by Bayes' theorem,
\begin{equation}
\label{eq:bayes}
P(\boldsymbol{\theta}|\mathbf{d})
= \frac{P(\mathbf{d}|\boldsymbol{\theta})\,P(\boldsymbol{\theta})}
{P(\mathbf{d})},
\end{equation}
where $P(\boldsymbol{\theta}|\mathbf{d})$ is the posterior probability 
distribution of the parameters, $P(\boldsymbol{\theta})$ is the prior, and
\begin{equation}
\label{eq:evidence}
P(\mathbf{d}) \equiv \mathcal{Z}
= \int P(\mathbf{d}|\boldsymbol{\theta})\,P(\boldsymbol{\theta})\,
\mathrm{d}\boldsymbol{\theta}
\end{equation}
is the Bayesian evidence (or marginal likelihood), which acts as a normalization constant independent of $\boldsymbol{\theta}$.
For parameter estimation, inference is based on the posterior, while  for model selection one compares evidences.

Neglecting noise, the log-likelihood for a GW signal $h_1$ and 
template $h_2$ reduces to
\begin{equation}
\label{eq:logL}
\log \mathcal{L} \simeq -\frac{1}{2} \langle h_1 - h_2 \,|\, h_1 - h_2 
\rangle + \text{const},
\end{equation}
where the additive constant depends only on $S_n(f)$ and is independent 
of the waveform parameters $\boldsymbol{\theta}$, so it plays no role in 
parameter estimation.
Assuming that both waveforms have the same norm, 
$\langle h_1 | h_1 \rangle = \langle h_2 | h_2 \rangle = \mathrm{SNR}^2$,
which holds when the two signals differ only in phase and time of 
coalescence and not in amplitude, we expand the inner product:
\begin{align}
\label{eq:logLapprox}
\langle h_1 - h_2 \,|\, h_1 - h_2 \rangle 
    &= \langle h_1 | h_1 \rangle + \langle h_2 | h_2 \rangle 
       - 2\,\langle h_1 | h_2 \rangle \nonumber \\
    &= 2\,\mathrm{SNR}^2 - 2\,\langle h_1 | h_2 \rangle \nonumber \\
    &= 2\,\mathrm{SNR}^2\,\mathcal{M},
\end{align}
where in the last step we used the definition \eqref{eq:MM}, with 
$\langle h_1 | h_2\rangle$.
Thus, dropping the additive constant, the log-likelihood can be 
approximated as
\begin{equation}
\label{eq:logL_estimation}
\log \mathcal{L} \sim -\mathrm{SNR}^2\,\mathcal{M}.
\end{equation}
On the other hand, in the high-SNR regime, which is particularly 
relevant for long-lived LISA sources, the norm of the waveform 
difference $\delta h = h_1 - h_2$ can be approximated 
as \cite{Cutler_1994}
\begin{align}
\label{eq:delta_logL}
\Delta\log\mathcal{L} 
    &= \log\mathcal{L}(\boldsymbol{\theta}_1) 
     - \log\mathcal{L}(\boldsymbol{\theta}_2) \nonumber \\
    &\simeq \tfrac{1}{2}\langle\delta h\,|\,\delta h\rangle 
     = \tfrac{1}{2}|\delta h|^2.
\end{align}
A standard criterion for two signals to be observationally 
distinguishable is that the likelihood changes by an $\mathcal{O}(1)$ 
amount, i.e.
\begin{equation}
\label{eq:detectability_1}
|\delta h| \gtrsim 1.
\end{equation}
For nearby signals $(\mathcal{M}\ll 1)$, expanding to leading order 
in $\mathcal{M}$ gives
\begin{equation}
\label{eq:detectability_2}
|\delta h|^2 = 2\,\langle h\,|\,h\rangle\,\mathcal{M} 
= 2\,\mathrm{SNR}^2\,\mathcal{M}.
\end{equation}
Combining \eqref{eq:detectability_1} and \eqref{eq:detectability_2} 
immediately yields the detectability requirement
\begin{equation}
\label{eq:SNR_condition}
\mathrm{SNR} \gtrsim \frac{1}{\sqrt{2\mathcal{M}}}.
\end{equation}
This relation shows that smaller waveform mismatches require proportionally higher SNR to be distinguishable in the data.

\subsection{Post-Newtonian terms}

We have considered the case of an evolving GA with a radial profile, and have shown that the timescale for the halo formation is generally longer than the timescale for the binary evolution. Focusing on the region of parameter space $(f_a, m_{dm})$ which causes a higher mismatch with respect to the vacuum waveforms, by making the approximation $\rho_{GA} \sim \rho_{crit} = 16 f_a^2 m_{d m}^4 M_{BH}^2$, the accumulated orbital phase depending on frequency can be computed analytically in terms 
\begin{align}
\label{eq:tw}
t(f_{\text{GW}}) &= \int^{f^{max}_{\text{GW}}}_{f^{min}_{\text{GW}}} \mathrm{d} f^{\prime}_{\text{GW}} \frac{1}{f_{\text{GW}}}\notag =\\
&=\frac{f_{\text{GW}}^2 \pi \, }{12\, q\, \rho}{}_2F_1\left( \frac{6}{11},\, 1;\, \frac{17}{11};\, -\frac{8 f_{\text{GW}}^{11/3} M_{BH}^{5/3} \pi^{8/3}}{5 \rho} \right)
\end{align}
\begin{align}
\label{eq:phiw}
\phi(f_{\text{GW}}) &= \int_{f^{min}_{\text{GW}}}^{f^{max}_{\text{GW}}} \mathrm{d} f_{\text{GW}}^{\prime} \frac{2 \pi f_{\text{GW}}^{\prime}}{f_{\text{GW}}^{\prime}}\notag =\\
&=\frac{f_{\text{GW}}}{12 q\, \rho}f_{\text{GW}} \, {}_2F_1\left( \frac{3}{11},\, 1;\, \frac{14}{11};\, -\frac{8 f_{\text{GW}}^{11/3} m^{5/3} \pi^{8/3}}{5 \rho} \right)
\end{align}
where $\rho = \rho(f, t)$ is given by \eqref{eq:density_profile}. Although exact, the expressions in \eqref{eq:tw} and \eqref{eq:phiw} involve Gauss hypergeometric functions $_2F_1$ whose argument mixes contributions from all PN orders, making it impossible to read off the PN order of the DF correction directly from these results without. In the lines of \cite{CanevaSantoro:2023aol}, is possible to expand the integrand for $\rho_{GA}m \ll 1$ and use the stationary phase approximation. At leading post-Newtonian order and for the dominant $m=2$ harmonic, the Fourier transforms of the plus and cross polarizations read
\cite{Cutler_1994}
\begin{equation}
    \tilde{h}_{+,\times}(f_{GW}) = \mathcal{A}_{+,\times}(f_{GW})\, e^{i\phi_{+,\times}(f_{GW})},
\end{equation}
Focusing on the phase, this is given for both polarizations as
\begin{equation}
    \phi_{+} \approx 2 \pi f_{GW} t(f_{GW})-\varphi(f_{GW})-\frac{\pi}{4}
\end{equation}
\begin{equation}
    \phi_{\times}=\phi_{+}+\frac{\pi}{2}
\end{equation}
By and this approximation for the integrands in  \eqref{eq:tw}  and \eqref{eq:phiw} leads to the result for the phase
\begin{equation}
\begin{aligned}
\phi_{+} & \approx 2 \pi f t_{\mathrm{c}}-\varphi_{\mathrm{c}}-\frac{\pi}{4} \\
& +\frac{3}{128 \eta  \pi^{5/3}M^{5/3}f_{GW}^{5/3}}\left(1+\delta_{\mathrm{DF}} + \delta_{\text{ACC}}\right)
\end{aligned}
\end{equation}
where
\begin{equation}
    \delta_{\mathrm{DF}}\approx-\frac{25 \rho}{304\left(\pi^{8 / 3} \eta^2 M_{BH}^{5 / 3}f_{GW}^{11 / 3}\right)}.
\end{equation}
\begin{equation}
\delta_{\mathrm{ACC}} \approx -\frac{125\rho}{1824\,\pi^{8/3}\eta^3M_{BH}^{5/3}\, f_{GW}^{11/3}} 
\,
\end{equation}
as in \cite{CanevaSantoro:2023aol}. In order to identify the leading PN order for this term, we just have to recall that at 0PN, the GW phase evolution scales as $\dot f_{\mathrm{GW}}\propto f_{\mathrm{GW}}^{11/3}$, with PN parameter $x\sim(\pi M_{BH} f_{\mathrm{GW}})^{2/3}$. Since $\delta_{\mathrm{DF}}\propto f_{\mathrm{GW}}^{-11/3}\sim x^{-11/2}$, it corresponds to a $-11/2=-5.5$PN correction. In the limit $f_a \to \infty \;\Rightarrow\; \rho \to 0$, the standard vacuum frequency evolution is recovered. 

In order to identify the leading PN order for this term, we just have to recall that at 0PN, 
the GW phase evolution scales as $\dot f_{\mathrm{GW}}\propto f_{\mathrm{GW}}^{11/3}$, with 
PN parameter $x\sim(\pi M_{BH} f_{\mathrm{GW}})^{2/3}$. Since $\delta_{\mathrm{DF}}\propto 
f_{\mathrm{GW}}^{-11/3}\sim x^{-11/2}$, it corresponds to a $-11/2=-5.5$PN correction. Similarly, since $\delta_{\mathrm{ACC}}\propto v^{-11}\sim x^{-11/2}$, it also enters at $-5.5$PN order. In the limit $f_a \to\infty \;\Rightarrow\; \rho \to 0$, the standard vacuum frequency evolution is recovered.

Because these corrections enter at negative PN order, they are particularly relevant for the early inspiral. The distinct frequency scaling of the $-5.5$PN term prevents it from being absorbed into a redefinition of the chirp mass, and explains why we did not find any correlation in the parameter estimation for the DM parameters and the binary intrinsic parameters.

\bibliography{./bibliography}

@article{Chia:2022udn,
    author = "Chia, Horng Sheng and Doorman, Christoffel and Wernersson, Alexandra and Hinderer, Tanja and Nissanke, Samaya",
    title = "{Self-interacting gravitational atoms in the strong-gravity regime}",
    eprint = "2212.11948",
    archivePrefix = "arXiv",
    primaryClass = "gr-qc",
    doi = "10.1088/1475-7516/2023/04/018",
    journal = "JCAP",
    volume = "04",
    pages = "018",
    year = "2023"
}

@article{Abbott_2019,
   title={GWTC-1: A Gravitational-Wave Transient Catalog of Compact Binary Mergers Observed by LIGO and Virgo during the First and Second Observing Runs},
   volume={9},
   ISSN={2160-3308},
   url={http://dx.doi.org/10.1103/PhysRevX.9.031040},
   DOI={10.1103/physrevx.9.031040},
   number={3},
   journal={Physical Review X},
   publisher={American Physical Society (APS)},
   author={Abbott, B. P. and others},
   year={2019},
   month=sep }

@misc{amaroseoane2017laserinterferometerspaceantenna,
      title={Laser Interferometer Space Antenna}, 
      author={Pau Amaro-Seoane and others},
      year={2017},
      eprint={1702.00786},
      archivePrefix={arXiv},
      primaryClass={astro-ph.IM},
      url={https://arxiv.org/abs/1702.00786}, 
}

@misc{cole2022disksspikescloudsdistinguishing,
      title={Disks, spikes, and clouds: distinguishing environmental effects on BBH gravitational waveforms}, 
      author={Philippa S. Cole and Gianfranco Bertone and Adam Coogan and Daniele Gaggero and Theophanes Karydas and Bradley J. Kavanagh and Thomas F. M. Spieksma and Giovanni Maria Tomaselli},
      year={2022},
      eprint={2211.01362},
      archivePrefix={arXiv},
      primaryClass={gr-qc},
      url={https://arxiv.org/abs/2211.01362}, 
}

@misc{budker2023genericformationmechanismultralight,
      title={A Generic Formation Mechanism of Ultralight Dark Matter Solar Halos}, 
      author={Dmitry Budker and Joshua Eby and Marco Gorghetto and Minyuan Jiang and Gilad Perez},
      year={2023},
      eprint={2306.12477},
      archivePrefix={arXiv},
      primaryClass={hep-ph},
      url={https://arxiv.org/abs/2306.12477}, 
}

@article{Tomaselli_2023,
   title={Dynamical friction in gravitational atoms},
   volume={2023},
   ISSN={1475-7516},
   url={http://dx.doi.org/10.1088/1475-7516/2023/07/070},
   DOI={10.1088/1475-7516/2023/07/070},
   number={07},
   journal={Journal of Cosmology and Astroparticle Physics},
   publisher={IOP Publishing},
   author={Tomaselli, Giovanni Maria and Spieksma, Thomas F.M. and Bertone, Gianfranco},
   year={2023},
   month=jul, pages={070} }

@article{Ferreira_2021,
   title={Ultra-light dark matter},
   volume={29},
   ISSN={1432-0754},
   url={http://dx.doi.org/10.1007/s00159-021-00135-6},
   DOI={10.1007/s00159-021-00135-6},
   number={1},
   journal={The Astronomy and Astrophysics Review},
   publisher={Springer Science and Business Media LLC},
   author={Ferreira, Elisa G. M.},
   year={2021},
   month=sep }

@article{Kavanagh_2020,
   title={Detecting dark matter around black holes with gravitational waves: Effects of dark-matter dynamics on the gravitational waveform},
   volume={102},
   ISSN={2470-0029},
   url={http://dx.doi.org/10.1103/PhysRevD.102.083006},
   DOI={10.1103/physrevd.102.083006},
   number={8},
   journal={Physical Review D},
   publisher={American Physical Society (APS)},
   author={Kavanagh, Bradley J. and Nichols, David A. and Bertone, Gianfranco and Gaggero, Daniele},
   year={2020},
   month=oct }

@article{Bertone_2020,
   title={Gravitational wave probes of dark matter: challenges and opportunities},
   volume={3},
   ISSN={2666-9366},
   url={http://dx.doi.org/10.21468/SciPostPhysCore.3.2.007},
   DOI={10.21468/scipostphyscore.3.2.007},
   number={2},
   journal={SciPost Physics Core},
   publisher={Stichting SciPost},
   author={Bertone, Gianfranco and Croon, Djuna and Amin, Mustafa and Boddy, Kimberly K. and Kavanagh, Bradley and Mack, Katherine J. and Natarajan, Priyamvada and Opferkuch, Toby and Schutz, Katelin and Takhistov, Volodymyr and Weniger, Christoph and Yu, Tien-Tien},
   year={2020},
   month=oct }

@article{PhysRevLett.38.1440,
  title = {$\mathrm{CP}$ Conservation in the Presence of Pseudoparticles},
  author = {Peccei, R. D. and Quinn, Helen R.},
  journal = {Phys. Rev. Lett.},
  volume = {38},
  issue = {25},
  pages = {1440--1443},
  numpages = {0},
  year = {1977},
  month = {Jun},
  publisher = {American Physical Society},
  doi = {10.1103/PhysRevLett.38.1440},
  url = {https://link.aps.org/doi/10.1103/PhysRevLett.38.1440}
}

@article{PhysRevLett.40.223,
  title = {A New Light Boson?},
  author = {Weinberg, Steven},
  journal = {Phys. Rev. Lett.},
  volume = {40},
  issue = {4},
  pages = {223--226},
  numpages = {0},
  year = {1978},
  month = {Jan},
  publisher = {American Physical Society},
  doi = {10.1103/PhysRevLett.40.223},
  url = {https://link.aps.org/doi/10.1103/PhysRevLett.40.223}
}

@article{PhysRevLett.40.279,
  title = {Problem of Strong $P$ and $T$ Invariance in the Presence of Instantons},
  author = {Wilczek, F.},
  journal = {Phys. Rev. Lett.},
  volume = {40},
  issue = {5},
  pages = {279--282},
  numpages = {0},
  year = {1978},
  month = {Jan},
  publisher = {American Physical Society},
  doi = {10.1103/PhysRevLett.40.279},
  url = {https://link.aps.org/doi/10.1103/PhysRevLett.40.279}
}

@article{PhysRevD.95.043541,
  title = {Ultralight scalars as cosmological dark matter},
  author = {Hui, Lam and Ostriker, Jeremiah P. and Tremaine, Scott and Witten, Edward},
  journal = {Phys. Rev. D},
  volume = {95},
  issue = {4},
  pages = {043541},
  numpages = {32},
  year = {2017},
  month = {Feb},
  publisher = {American Physical Society},
  doi = {10.1103/PhysRevD.95.043541},
  url = {https://link.aps.org/doi/10.1103/PhysRevD.95.043541}
}

@article{Svrcek_2006,
   title={Axions in string theory},
   volume={2006},
   ISSN={1029-8479},
   url={http://dx.doi.org/10.1088/1126-6708/2006/06/051},
   DOI={10.1088/1126-6708/2006/06/051},
   number={06},
   journal={Journal of High Energy Physics},
   publisher={Springer Science and Business Media LLC},
   author={Svrcek, Peter and Witten, Edward},
   year={2006},
   month=jun, pages={051–051} }

@article{Liebling_2023,
   title={Dynamical boson stars},
   volume={26},
   ISSN={1433-8351},
   url={http://dx.doi.org/10.1007/s41114-023-00043-4},
   DOI={10.1007/s41114-023-00043-4},
   number={1},
   journal={Living Reviews in Relativity},
   publisher={Springer Science and Business Media LLC},
   author={Liebling, Steven L. and Palenzuela, Carlos},
   year={2023},
   month=feb }

@article{Bezares_2017,
   title={Final fate of compact boson star mergers},
   volume={95},
   ISSN={2470-0029},
   url={http://dx.doi.org/10.1103/PhysRevD.95.124005},
   DOI={10.1103/physrevd.95.124005},
   number={12},
   journal={Physical Review D},
   publisher={American Physical Society (APS)},
   author={Bezares, Miguel and Palenzuela, Carlos and Bona, Carles},
   year={2017},
   month=jun }

@article{Chavanis:2019bnu,
    author = "Chavanis, Pierre-Henri",
    title = "{Mass-radius relation of self-gravitating Bose-Einstein condensates with a central black hole}",
    eprint = "1909.04709",
    archivePrefix = "arXiv",
    primaryClass = "gr-qc",
    doi = "10.1140/epjp/i2019-12734-7",
    journal = "Eur. Phys. J. Plus",
    volume = "134",
    number = "7",
    pages = "352",
    year = "2019"
}

@article{Chatziioannou:2017tdw,
    author = "Chatziioannou, Katerina and Klein, Antoine and Yunes, Nicol{\'a}s and Cornish, Neil",
    title = "{Constructing Gravitational Waves from Generic Spin-Precessing Compact Binary Inspirals}",
    eprint = "1703.03967",
    archivePrefix = "arXiv",
    primaryClass = "gr-qc",
    doi = "10.1103/PhysRevD.95.104004",
    journal = "Phys. Rev. D",
    volume = "95",
    number = "10",
    pages = "104004",
    year = "2017"
}

@article{JimenezForteza:2018rwr,
    author = "Jim{\'e}nez Forteza, Xisco and Abdelsalhin, Tiziano and Pani, Paolo and Gualtieri, Leonardo",
    title = "{Impact of high-order tidal terms on binary neutron-star waveforms}",
    eprint = "1807.08016",
    archivePrefix = "arXiv",
    primaryClass = "gr-qc",
    doi = "10.1103/PhysRevD.98.124014",
    journal = "Phys. Rev. D",
    volume = "98",
    number = "12",
    pages = "124014",
    year = "2018"
}

@article{Bezares_2018,
   title={Gravitational waves from dark boson star binary mergers},
   volume={35},
   ISSN={1361-6382},
   url={http://dx.doi.org/10.1088/1361-6382/aae87c},
   DOI={10.1088/1361-6382/aae87c},
   number={23},
   journal={Classical and Quantum Gravity},
   publisher={IOP Publishing},
   author={Bezares, Miguel and Palenzuela, Carlos},
   year={2018},
   month=nov, pages={234002} }

@article{Press:1972zz,
    author = "Press, William H. and Teukolsky, Saul A.",
    title = "{Floating Orbits, Superradiant Scattering and the Black-hole Bomb}",
    doi = "10.1038/238211a0",
    journal = "Nature",
    volume = "238",
    pages = "211--212",
    year = "1972"
}

@book{Brito_2020,
   title={Superradiance: New Frontiers in Black Hole Physics},
   ISBN={9783030466220},
   ISSN={1616-6361},
   url={http://dx.doi.org/10.1007/978-3-030-46622-0},
   DOI={10.1007/978-3-030-46622-0},
   journal={Lecture Notes in Physics},
   publisher={Springer International Publishing},
   author={Brito, Richard and Cardoso, Vitor and Pani, Paolo},
   year={2020} }

@article{Eda_2013,
   title={New Probe of Dark-Matter Properties: Gravitational Waves from an Intermediate-Mass Black Hole Embedded in a Dark-Matter Minispike},
   volume={110},
   ISSN={1079-7114},
   url={http://dx.doi.org/10.1103/PhysRevLett.110.221101},
   DOI={10.1103/physrevlett.110.221101},
   number={22},
   journal={Physical Review Letters},
   publisher={American Physical Society (APS)},
   author={Eda, Kazunari and Itoh, Yousuke and Kuroyanagi, Sachiko and Silk, Joseph},
   year={2013},
   month=may }

@article{Eda_2015,
   title={Gravitational waves as a probe of dark matter minispikes},
   volume={91},
   ISSN={1550-2368},
   url={http://dx.doi.org/10.1103/PhysRevD.91.044045},
   DOI={10.1103/physrevd.91.044045},
   number={4},
   journal={Physical Review D},
   publisher={American Physical Society (APS)},
   author={Eda, Kazunari and Itoh, Yousuke and Kuroyanagi, Sachiko and Silk, Joseph},
   year={2015},
   month=feb }

@article{Zhao_2005,
   title={Dark Minihalos with Intermediate Mass Black Holes},
   volume={95},
   ISSN={1079-7114},
   url={http://dx.doi.org/10.1103/PhysRevLett.95.011301},
   DOI={10.1103/physrevlett.95.011301},
   number={1},
   journal={Physical Review Letters},
   publisher={American Physical Society (APS)},
   author={Zhao, HongSheng and Silk, Joseph},
   year={2005},
   month=jun }

@article{Barausse_2015,
   title={Environmental Effects for Gravitational-wave Astrophysics},
   volume={610},
   ISSN={1742-6596},
   url={http://dx.doi.org/10.1088/1742-6596/610/1/012044},
   DOI={10.1088/1742-6596/610/1/012044},
   journal={Journal of Physics: Conference Series},
   publisher={IOP Publishing},
   author={Barausse, Enrico and Cardoso, Vitor and Pani, Paolo},
   year={2015},
   month=may, pages={012044} }

@article{Dolan_2007,
   title={Instability of the massive Klein-Gordon field on the Kerr spacetime},
   volume={76},
   ISSN={1550-2368},
   url={http://dx.doi.org/10.1103/PhysRevD.76.084001},
   DOI={10.1103/physrevd.76.084001},
   number={8},
   journal={Physical Review D},
   publisher={American Physical Society (APS)},
   author={Dolan, Sam R.},
   year={2007},
   month=oct }

@article{Chandra_DF,
   title={Dynamical Friction. I. General Considerations: the Coefficient of Dynamical Friction},
   volume={97},
   url={https://ui.adsabs.harvard.edu/abs/1943ApJ....97..255C/abstract},
   DOI={10.1086/144517},
   journal={Astrophysical Journal},
   publisher={American Astronomical Society (AAS)},
   author={Chandrasekhar, S.},
   year={1943}}

@article{Hui_2017,
   title={Ultralight scalars as cosmological dark matter},
   volume={95},
   ISSN={2470-0029},
   url={http://dx.doi.org/10.1103/PhysRevD.95.043541},
   DOI={10.1103/physrevd.95.043541},
   number={4},
   journal={Physical Review D},
   publisher={American Physical Society (APS)},
   author={Hui, Lam and Ostriker, Jeremiah P. and Tremaine, Scott and Witten, Edward},
   year={2017},
   month=feb }

@article{Bondi:1952ni,
    author = "Bondi, H.",
    title = "{On spherically symmetrical accretion}",
    doi = "10.1093/mnras/112.2.195",
    journal = "Mon. Not. Roy. Astron. Soc.",
    volume = "112",
    pages = "195",
    year = "1952"
}

@article{Bondi:1944rnk,
    author = "Bondi, H. and Hoyle, F.",
    title = "{On the Mechanism of Accretion by Stars}",
    doi = "10.1093/mnras/104.5.273",
    journal = "Mon. Not. Roy. Astron. Soc.",
    volume = "104",
    number = "5",
    pages = "273--282",
    year = "1944"
}

@article{CanevaSantoro:2023aol,
    author = "Caneva Santoro, Giada and Roy, Soumen and Vicente, Rodrigo and Haney, Maria and Piccinni, Ornella Juliana and Del Pozzo, Walter and Martinez, Mario",
    title = "{First Constraints on Compact Binary Environments from LIGO-Virgo Data}",
    eprint = "2309.05061",
    archivePrefix = "arXiv",
    primaryClass = "gr-qc",
    reportNumber = "LIGO DCC P2300301",
    doi = "10.1103/PhysRevLett.132.251401",
    journal = "Phys. Rev. Lett.",
    volume = "132",
    number = "25",
    pages = "251401",
    year = "2024"
}

@article{Degollado_2018,
   title={Effective stability against superradiance of Kerr black holes with synchronised hair},
   volume={781},
   ISSN={0370-2693},
   url={http://dx.doi.org/10.1016/j.physletb.2018.04.052},
   DOI={10.1016/j.physletb.2018.04.052},
   journal={Physics Letters B},
   publisher={Elsevier BV},
   author={Degollado, Juan Carlos and Herdeiro, Carlos A.R. and Radu, Eugen},
   year={2018},
   month=jun, pages={651–655} }

@article{Pere_iguez_2024,
   title={Superradiant instability of magnetic black holes},
   volume={110},
   ISSN={2470-0029},
   url={http://dx.doi.org/10.1103/PhysRevD.110.104001},
   DOI={10.1103/physrevd.110.104001},
   number={10},
   journal={Physical Review D},
   publisher={American Physical Society (APS)},
   author={Pereñiguez, David and de Amicis, Marina and Brito, Richard and Macedo, Rodrigo Panosso},
   year={2024},
   month=nov }

@article{Gross:1961kqh,
    author = "Gross, E. P.",
    title = "{Structure of a quantized vortex in boson systems}",
    doi = "10.1007/BF02731494",
    journal = "Nuovo Cim.",
    volume = "20",
    number = "3",
    pages = "454--477",
    year = "1961"
}

@article{Pitaevskii_1996,
   title={Dynamics of collapse of a confined Bose gas},
   volume={221},
   ISSN={0375-9601},
   url={http://dx.doi.org/10.1016/0375-9601(96)00538-5},
   DOI={10.1016/0375-9601(96)00538-5},
   number={1–2},
   journal={Physics Letters A},
   publisher={Elsevier BV},
   author={Pitaevskii, L.P.},
   year={1996},
   month=sep, pages={14–18} }

@article{Arcadi_2025,
   title={The waning of the WIMP: endgame?},
   volume={85},
   ISSN={1434-6052},
   url={http://dx.doi.org/10.1140/epjc/s10052-024-13672-y},
   DOI={10.1140/epjc/s10052-024-13672-y},
   number={2},
   journal={The European Physical Journal C},
   publisher={Springer Science and Business Media LLC},
   author={Arcadi, Giorgio and Cabo-Almeida, David and Dutra, Maíra and Ghosh, Pradipta and Lindner, Manfred and Mambrini, Yann and Neto, Jacinto P. and Pierre, Mathias and Profumo, Stefano and Queiroz, Farinaldo S.},
   year={2025},
   month=feb }

@article{Navarro_1997,
   title={A Universal Density Profile from Hierarchical Clustering},
   volume={490},
   ISSN={1538-4357},
   url={http://dx.doi.org/10.1086/304888},
   DOI={10.1086/304888},
   number={2},
   journal={The Astrophysical Journal},
   publisher={American Astronomical Society},
   author={Navarro, Julio F. and Frenk, Carlos S. and White, Simon D. M.},
   year={1997},
   month=dec, pages={493–508} }

@article{Madau_2001,
   title={Massive Black Holes as Population III Remnants},
   volume={551},
   ISSN={0004-637X},
   url={http://dx.doi.org/10.1086/319848},
   DOI={10.1086/319848},
   number={1},
   journal={The Astrophysical Journal},
   publisher={American Astronomical Society},
   author={Madau, Piero and Rees, Martin J.},
   year={2001},
   month=apr, pages={L27–L30} }

@article{Heger_2003,
   title={How Massive Single Stars End Their Life},
   volume={591},
   ISSN={1538-4357},
   url={http://dx.doi.org/10.1086/375341},
   DOI={10.1086/375341},
   number={1},
   journal={The Astrophysical Journal},
   publisher={American Astronomical Society},
   author={Heger, A. and Fryer, C. L. and Woosley, S. E. and Langer, N. and Hartmann, D. H.},
   year={2003},
   month=jul, pages={288–300} }

@article{Portegies_Zwart_2002,
   title={The Runaway Growth of Intermediate‐Mass Black Holes in Dense Star Clusters},
   volume={576},
   ISSN={1538-4357},
   url={http://dx.doi.org/10.1086/341798},
   DOI={10.1086/341798},
   number={2},
   journal={The Astrophysical Journal},
   publisher={American Astronomical Society},
   author={Portegies Zwart, Simon F. and McMillan, Stephen L. W.},
   year={2002},
   month=sep, pages={899–907} }

@article{Loeb_1994,
   title={Collapse of primordial gas clouds and the formation of quasar black holes},
   volume={432},
   ISSN={1538-4357},
   url={http://dx.doi.org/10.1086/174548},
   DOI={10.1086/174548},
   journal={The Astrophysical Journal},
   publisher={American Astronomical Society},
   author={Loeb, Abraham and Rasio, Frederic A.},
   year={1994},
   month=sep, pages={52} }

@article{Aurrekoetxea_2024,
   title={Self-interacting scalar dark matter around binary black holes},
   volume={110},
   ISSN={2470-0029},
   url={http://dx.doi.org/10.1103/PhysRevD.110.083011},
   DOI={10.1103/physrevd.110.083011},
   number={8},
   journal={Physical Review D},
   publisher={American Physical Society (APS)},
   author={Aurrekoetxea, Josu C. and Marsden, James and Clough, Katy and Ferreira, Pedro G.},
   year={2024},
   month=oct }

@article{Barausse_2014,
   title={Can environmental effects spoil precision gravitational-wave astrophysics?},
   volume={89},
   ISSN={1550-2368},
   url={http://dx.doi.org/10.1103/PhysRevD.89.104059},
   DOI={10.1103/physrevd.89.104059},
   number={10},
   journal={Physical Review D},
   publisher={American Physical Society (APS)},
   author={Barausse, Enrico and Cardoso, Vitor and Pani, Paolo},
   year={2014},
   month=may }

@article{Boudon_2024,
   title={Gravitational waves from binary black holes in a self-interacting scalar dark matter cloud},
   volume={109},
   ISSN={2470-0029},
   url={http://dx.doi.org/10.1103/PhysRevD.109.043504},
   DOI={10.1103/physrevd.109.043504},
   number={4},
   journal={Physical Review D},
   publisher={American Physical Society (APS)},
   author={Boudon, Alexis and Brax, Philippe and Valageas, Patrick and Wong, Leong Khim},
   year={2024},
   month=feb }

@book{Maggiore2008,
  author    = {Michele Maggiore},
  title     = {Gravitational Waves: Volume 1: Theory and Experiments},
  publisher = {Oxford University Press},
  year      = {2008},
  address   = {Oxford, UK},
  isbn      = {978-0-19-857074-5}
}

@article{Robson_2019,
   title={The construction and use of LISA sensitivity curves},
   volume={36},
   ISSN={1361-6382},
   url={http://dx.doi.org/10.1088/1361-6382/ab1101},
   DOI={10.1088/1361-6382/ab1101},
   number={10},
   journal={Classical and Quantum Gravity},
   publisher={IOP Publishing},
   author={Robson, Travis and Cornish, Neil J and Liu, Chang},
   year={2019},
   month=apr, pages={105011} }

@article{Cutler_1994,
   title={Gravitational waves from merging compact binaries: How accurately can one extract the binary’s parameters from the inspiral waveform?},
   volume={49},
   ISSN={0556-2821},
   url={http://dx.doi.org/10.1103/PhysRevD.49.2658},
   DOI={10.1103/physrevd.49.2658},
   number={6},
   journal={Physical Review D},
   publisher={American Physical Society (APS)},
   author={Cutler, Curt and Flanagan, Éanna E.},
   year={1994},
   month=mar, pages={2658–2697} }

@article{Cole_2023,
   title={Distinguishing environmental effects on binary black hole gravitational waveforms},
   volume={7},
   ISSN={2397-3366},
   url={http://dx.doi.org/10.1038/s41550-023-01990-2},
   DOI={10.1038/s41550-023-01990-2},
   number={8},
   journal={Nature Astronomy},
   publisher={Springer Science and Business Media LLC},
   author={Cole, Philippa S. and Bertone, Gianfranco and Coogan, Adam and Gaggero, Daniele and Karydas, Theophanes and Kavanagh, Bradley J. and Spieksma, Thomas F. M. and Tomaselli, Giovanni Maria},
   year={2023},
   month=jun, pages={943–950} }

@article{Baumann_2022,
   title={Ionization of gravitational atoms},
   volume={105},
   ISSN={2470-0029},
   url={http://dx.doi.org/10.1103/PhysRevD.105.115036},
   DOI={10.1103/physrevd.105.115036},
   number={11},
   journal={Physical Review D},
   publisher={American Physical Society (APS)},
   author={Baumann, Daniel and Bertone, Gianfranco and Stout, John and Tomaselli, Giovanni Maria},
   year={2022},
   month=jun }

@article{Arvanitaki_2010,
   title={String axiverse},
   volume={81},
   ISSN={1550-2368},
   url={http://dx.doi.org/10.1103/PhysRevD.81.123530},
   DOI={10.1103/physrevd.81.123530},
   number={12},
   journal={Physical Review D},
   publisher={American Physical Society (APS)},
   author={Arvanitaki, Asimina and Dimopoulos, Savas and Dubovsky, Sergei and Kaloper, Nemanja and March-Russell, John},
   year={2010},
   month=jun }

@misc{ou2023darkmatterprofilemilky,
      title={The dark matter profile of the Milky Way inferred from its circular velocity curve}, 
      author={Xiaowei Ou and Anna-Christina Eilers and Lina Necib and Anna Frebel},
      year={2023},
      eprint={2303.12838},
      archivePrefix={arXiv},
      primaryClass={astro-ph.GA},
      url={https://arxiv.org/abs/2303.12838}, 
}

@article{Gondolo_1999,
   title={Dark Matter Annihilation at the Galactic Center},
   volume={83},
   ISSN={1079-7114},
   url={http://dx.doi.org/10.1103/PhysRevLett.83.1719},
   DOI={10.1103/physrevlett.83.1719},
   number={9},
   journal={Physical Review Letters},
   publisher={American Physical Society (APS)},
   author={Gondolo, Paolo and Silk, Joseph},
   year={1999},
   month=aug, pages={1719–1722} }

@ARTICLE{1990ApJ...356..359H,
       author = {{Hernquist}, Lars},
        title = "{An Analytical Model for Spherical Galaxies and Bulges}",
      journal = {\apj},
         year = 1990,
        month = jun,
       volume = {356},
        pages = {359},
          doi = {10.1086/168845},
       adsurl = {https://ui.adsabs.harvard.edu/abs/1990ApJ...356..359H}
}

@ARTICLE{1965TrAlm...5...87E,
       author = {{Einasto}, J.},
        title = "{On the Construction of a Composite Model for the Galaxy and on the Determination of the System of Galactic Parameters}",
      journal = {Trudy Astrofizicheskogo Instituta Alma-Ata},
         year = 1965,
        month = jan,
       volume = {5},
        pages = {87-100},
       adsurl = {https://ui.adsabs.harvard.edu/abs/1965TrAlm...5...87E}
}

@article{Ir_i__2017,
   title={First Constraints on Fuzzy Dark Matter from Lyman-
<mml:math xmlns:mml="http://www.w3.org/1998/Math/MathML" display="inline"><mml:mi>α</mml:mi></mml:math>
 Forest Data and Hydrodynamical Simulations},
   volume={119},
   ISSN={1079-7114},
   url={http://dx.doi.org/10.1103/PhysRevLett.119.031302},
   DOI={10.1103/physrevlett.119.031302},
   number={3},
   journal={Physical Review Letters},
   publisher={American Physical Society (APS)},
   author={Iršič, Vid and Viel, Matteo and Haehnelt, Martin G. and Bolton, James S. and Becker, George D.},
   year={2017},
   month=jul }

@article{Marsh_2016,
   title={Axion cosmology},
   volume={643},
   ISSN={0370-1573},
   url={http://dx.doi.org/10.1016/j.physrep.2016.06.005},
   DOI={10.1016/j.physrep.2016.06.005},
   journal={Physics Reports},
   publisher={Elsevier BV},
   author={Marsh, David J.E.},
   year={2016},
   month=jul, pages={1–79} }

@inproceedings{Raffelt, series={COSMICWISPers},
   title={Astrophysical Axion Bounds: The 2024 Edition},
   url={http://dx.doi.org/10.22323/1.454.0041},
   DOI={10.22323/1.454.0041},
   booktitle={Proceedings of 1st General Meeting and 1st Training School of the COST Action COSMIC WSIPers — PoS(COSMICWISPers)},
   publisher={Sissa Medialab},
   author={Raffelt, G. and Caputo, A.},
   year={2024},
   month=mar, pages={041},
   collection={COSMICWISPers} }

@misc{banik2025bosonstarshostingblack,
      title={Boson Stars Hosting Black Holes}, 
      author={Amitayus Banik and Jeong Han Kim and Xing-Yu Yang},
      year={2025},
      eprint={2511.03788},
      archivePrefix={arXiv},
      primaryClass={gr-qc},
      url={https://arxiv.org/abs/2511.03788}, 
}

@misc{LISA_waveforms_2023,
      title={Waveform Modelling for the Laser Interferometer Space Antenna}, 
      author={LISA Consortium Waveform Working Group},
      year={2023},
      eprint={2311.01300},
      archivePrefix={arXiv},
      primaryClass={gr-qc},
      url={https://arxiv.org/abs/2311.01300}, 
}

@misc{blanchet2024,
      title={Post-Newtonian Theory for Gravitational Waves}, 
      author={Luc Blanchet},
      year={2024},
      eprint={1310.1528},
      archivePrefix={arXiv},
      primaryClass={gr-qc},
      url={https://arxiv.org/abs/1310.1528}, 
}

@article{Roy_2026,
   title={Scalar Fields around Black Hole Binaries in LIGO-Virgo-KAGRA},
   volume={136},
   ISSN={1079-7114},
   url={http://dx.doi.org/10.1103/fv9z-zkxx},
   DOI={10.1103/fv9z-zkxx},
   number={19},
   journal={Physical Review Letters},
   publisher={American Physical Society (APS)},
   author={Roy, Soumen and Vicente, Rodrigo and Aurrekoetxea, Josu C. and Clough, Katy and Ferreira, Pedro G.},
   year={2026},
   month=May }

@misc{colpi2024lisadefinitionstudyreport,
      title={LISA Definition Study Report}, 
      author={Monica Colpi and others},
      year={2024},
      eprint={2402.07571},
      archivePrefix={arXiv},
      primaryClass={astro-ph.CO},
      url={https://arxiv.org/abs/2402.07571}, 
}

@article{Abac_2026,
   title={The Science of the Einstein Telescope},
   volume={2026},
   ISSN={1475-7516},
   url={http://dx.doi.org/10.1088/1475-7516/2026/03/081},
   DOI={10.1088/1475-7516/2026/03/081},
   number={03},
   journal={Journal of Cosmology and Astroparticle Physics},
   publisher={IOP Publishing},
   author={Abac, Adrian and others},
   year={2026},
   month=Mar,
   pages={081}
}

@article{Traykova_2023,
   title={Relativistic drag forces on black holes from scalar dark matter clouds of all sizes},
   volume={108},
   ISSN={2470-0029},
   url={http://dx.doi.org/10.1103/PhysRevD.108.L121502},
   DOI={10.1103/physrevd.108.l121502},
   number={12},
   journal={Physical Review D},
   publisher={American Physical Society (APS)},
   author={Traykova, Dina and Vicente, Rodrigo and Clough, Katy and Helfer, Thomas and Berti, Emanuele and Ferreira, Pedro G. and Hui, Lam},
   year={2023},
   month=Dec }

@article{Abbott_2021,
   title={GWTC-2: Compact Binary Coalescences Observed by LIGO and Virgo during the First Half of the Third Observing Run},
   volume={11},
   ISSN={2160-3308},
   url={http://dx.doi.org/10.1103/PhysRevX.11.021053},
   DOI={10.1103/physrevx.11.021053},
   number={2},
   journal={Physical Review X},
   publisher={American Physical Society (APS)},
   author={Abbott, R. and others},
   year={2021},
   month=June}

@article{Abbott_2023,
   title={GWTC-3: Compact Binary Coalescences Observed by LIGO and Virgo during the Second Part of the Third Observing Run},
   volume={13},
   ISSN={2160-3308},
   url={http://dx.doi.org/10.1103/PhysRevX.13.041039},
   DOI={10.1103/physrevx.13.041039},
   number={4},
   journal={Physical Review X},
   publisher={American Physical Society (APS)},
   author={Abbott, R. and others},
   year={2023},
   month=Dec }

@misc{theligoscientificcollaboration2025gwtc40updatinggravitationalwavetransient,
      title={GWTC-4.0: Updating the Gravitational-Wave Transient Catalog with Observations from the First Part of the Fourth LIGO-Virgo-KAGRA Observing Run}, 
      author={The LIGO Scientific Collaboration and the Virgo Collaboration and the KAGRA Collaborationo},
      year={2025},
      eprint={2508.18082},
      archivePrefix={arXiv},
      primaryClass={gr-qc},
      url={https://arxiv.org/abs/2508.18082}, 
}

@article{Arun_2022,
   title={New horizons for fundamental physics with LISA},
   volume={25},
   ISSN={1433-8351},
   url={http://dx.doi.org/10.1007/s41114-022-00036-9},
   DOI={10.1007/s41114-022-00036-9},
   number={1},
   journal={Living Reviews in Relativity},
   publisher={Springer Science and Business Media LLC},
   author={Arun, K. G. and others},
   year={2022},
   month=June }

@article{Cardoso_2018,
   title={Constraining the mass of dark photons and axion-like particles through black-hole superradiance},
   volume={2018},
   ISSN={1475-7516},
   url={http://dx.doi.org/10.1088/1475-7516/2018/03/043},
   DOI={10.1088/1475-7516/2018/03/043},
   number={03},
   journal={Journal of Cosmology and Astroparticle Physics},
   publisher={IOP Publishing},
   author={Cardoso, Vitor and Dias, Óscar J.C. and Hartnett, Gavin S. and Middleton, Matthew and Pani, Paolo and Santos, Jorge E.},
   year={2018},
   month=Mar, pages={043–043} }

@article{Brito_2017,
   title={Gravitational wave searches for ultralight bosons with LIGO and LISA},
   volume={96},
   ISSN={2470-0029},
   url={http://dx.doi.org/10.1103/PhysRevD.96.064050},
   DOI={10.1103/physrevd.96.064050},
   number={6},
   journal={Physical Review D},
   publisher={American Physical Society (APS)},
   author={Brito, Richard and Ghosh, Shrobana and Barausse, Enrico and Berti, Emanuele and Cardoso, Vitor and Dvorkin, Irina and Klein, Antoine and Pani, Paolo},
   year={2017},
   month=Sept }

@article{Macedo_2013,
   title={INTO THE LAIR: GRAVITATIONAL-WAVE SIGNATURES OF DARK MATTER},
   volume={774},
   ISSN={1538-4357},
   url={http://dx.doi.org/10.1088/0004-637X/774/1/48},
   DOI={10.1088/0004-637x/774/1/48},
   number={1},
   journal={The Astrophysical Journal},
   publisher={American Astronomical Society},
   author={Macedo, Caio F. B. and Pani, Paolo and Cardoso, Vitor and Crispino, Luís C. B.},
   year={2013},
   month=Aug, pages={48} }

@article{Duque_2024,
   title={Extreme-Mass-Ratio Inspirals in Ultralight Dark Matter},
   volume={133},
   ISSN={1079-7114},
   url={http://dx.doi.org/10.1103/PhysRevLett.133.121404},
   DOI={10.1103/physrevlett.133.121404},
   number={12},
   journal={Physical Review Letters},
   publisher={American Physical Society (APS)},
   author={Duque, Francisco and Macedo, Caio F. B. and Vicente, Rodrigo and Cardoso, Vitor},
   year={2024},
   month=Sept }

@article{Yunes:2013dva,
    author = "Yunes, Nicol{\'a}s and Siemens, Xavier",
    title = "{Gravitational-Wave Tests of General Relativity with Ground-Based Detectors and Pulsar Timing-Arrays}",
    eprint = "1304.3473",
    archivePrefix = "arXiv",
    primaryClass = "gr-qc",
    doi = "10.12942/lrr-2013-9",
    journal = "Living Rev. Rel.",
    volume = "16",
    pages = "9",
    year = "2013"
}

@article{Askar_2019,
   title={Black holes, gravitational waves and fundamental physics: a roadmap},
   volume={36},
   ISSN={1361-6382},
   url={http://dx.doi.org/10.1088/1361-6382/ab0587},
   DOI={10.1088/1361-6382/ab0587},
   number={14},
   journal={Classical and Quantum Gravity},
   publisher={IOP Publishing},
   author={Askar, Abbas and others},
   year={2019},
   month=June, pages={143001} }

@article{Sasaki_2018,
   title={Primordial black holes—perspectives in gravitational wave astronomy},
   volume={35},
   ISSN={1361-6382},
   url={http://dx.doi.org/10.1088/1361-6382/aaa7b4},
   DOI={10.1088/1361-6382/aaa7b4},
   number={6},
   journal={Classical and Quantum Gravity},
   publisher={IOP Publishing},
   author={Sasaki, Misao and Suyama, Teruaki and Tanaka, Takahiro and Yokoyama, Shuichiro},
   year={2018},
   month=Feb, pages={063001} }

@article{Bartolo_2019,
   title={Testing primordial black holes as dark matter with LISA},
   volume={99},
   ISSN={2470-0029},
   url={http://dx.doi.org/10.1103/PhysRevD.99.103521},
   DOI={10.1103/physrevd.99.103521},
   number={10},
   journal={Physical Review D},
   publisher={American Physical Society (APS)},
   author={Bartolo, N. and De Luca, V. and Franciolini, G. and Peloso, M. and Racco, D. and Riotto, A.},
   year={2019},
   month=May}

@article{Maggiore_2020,
   title={Science case for the Einstein telescope},
   volume={2020},
   ISSN={1475-7516},
   url={http://dx.doi.org/10.1088/1475-7516/2020/03/050},
   DOI={10.1088/1475-7516/2020/03/050},
   number={03},
   journal={Journal of Cosmology and Astroparticle Physics},
   publisher={IOP Publishing},
   author={Maggiore, Michele and others},
   year={2020},
   month=Mar, pages={050–050} }

@article{Branchesi_2023,
   title={Science with the Einstein Telescope: a comparison of different designs},
   volume={2023},
   ISSN={1475-7516},
   url={http://dx.doi.org/10.1088/1475-7516/2023/07/068},
   DOI={10.1088/1475-7516/2023/07/068},
   number={07},
   journal={Journal of Cosmology and Astroparticle Physics},
   publisher={IOP Publishing},
   author={Branchesi, Marica and others},
   year={2023},
   month=July, pages={068} }

@article{Cardoso_2016,
   title={Gravitational-wave signatures of exotic compact objects and of quantum corrections at the horizon scale},
   volume={94},
   ISSN={2470-0029},
   url={http://dx.doi.org/10.1103/PhysRevD.94.084031},
   DOI={10.1103/physrevd.94.084031},
   number={8},
   journal={Physical Review D},
   publisher={American Physical Society (APS)},
   author={Cardoso, Vitor and Hopper, Seth and Macedo, Caio F. B. and Palenzuela, Carlos and Pani, Paolo},
   year={2016},
   month=Oct }

@misc{berti2025blackholespectroscopytheory,
      title={Black hole spectroscopy: from theory to experiment}, 
      author={Emanuele Berti and others},
      year={2025},
      eprint={2505.23895},
      archivePrefix={arXiv},
      primaryClass={gr-qc},
      url={https://arxiv.org/abs/2505.23895}, 
}
\end{document}